\documentclass[aps,pre,reprint,amsmath,amssymb,nofootinbib]{revtex4-2}
\usepackage[T1]{fontenc}
\usepackage[utf8]{inputenc}
\usepackage{graphicx,booktabs,bm,hyperref,tikz}
\usetikzlibrary{arrows.meta,positioning}
\hypersetup{colorlinks=true,citecolor=blue,linkcolor=blue,urlcolor=blue,
 pdftitle={Modulation equations and coherence dynamics in a qubit chain coupled to a circulating fluxon},
 pdfauthor={O. M. Kiselev}}
\newcommand{\dd}{\mathrm{d}}
\newcommand{\Tr}{\operatorname{Tr}}
\newcommand{\sech}{\operatorname{sech}}
\newcommand{\hstar}{\hbar_*}
\newcommand{\hc}{\mathrm{H.c.}}
\begin{document}
\title{Modulation equations and coherence dynamics in a qubit chain coupled to a circulating fluxon}
\author{O. M. Kiselev}
\affiliation{Innopolis University, Innopolis, Russia}
\date{November 23, 2026}
\begin{abstract}
We derive modulation equations for the joint quantum amplitudes of a circulating fluxon and an exchange-coupled qubit chain whose endpoint interacts with the field. A unitary passage operator combines the translational packet, vacuum dressing, and discrete oscillatory modes. A resonant normal form separates the slow amplitude evolution from nonresonant corrections. Ordered emission and return amplitudes generate a second-order correction that changes the population and coherence of an internal qubit. The next ground-band coefficient gives compensating diagonal and near-resonant off-diagonal contributions to the accumulated error. We bound the state error a posteriori. At a velocity-selected resonance, direct propagation over 81 passages gives a mode contribution of order $6\times10^{-8}$; the cubic ground-band approximation agrees with an expanded full calculation to $1.5\times10^{-9}$ in the tested observables. The equations describe how the source profile and field memory affect coherence inside the chain.
\end{abstract}
\maketitle

\section{Introduction}
A Josephson fluxon carries a localized phase winding and magnetic field along an extended junction. In an annular junction it can repeatedly encounter an inductively coupled flux qubit. Soliton-based readout was proposed in Ref.~\cite{fedorov2007}; coupling a moving fluxon to a superconducting qubit and detecting its spectrum through the fluxon oscillation frequency were demonstrated in Ref.~\cite{fedorov2014}. The field produced during one passage can affect subsequent encounters. The dynamics of an annular junction also depends on the coupling between the circulating fluxon and its oscillatory modes~\cite{ustinov1999}.

Soliton-mediated qubit manipulation has been studied for magnetic signals propagating through Heisenberg chains. A passing magnetic soliton can rotate a locally coupled qubit, with the response determined by the soliton profile and coupling geometry~\cite{cuccoli2015}. Bright- and dark-soliton control can also be described by a two-level Hamiltonian whose matrix elements depend on the soliton parameters~\cite{varbev2022}. In superconducting circuits, a recent soliton digital-to-analog converter uses a soliton held in a load loop to apply a controllable flux and implement transmon phase gates~\cite{reitz2026}.

Exchange transmits a local perturbation into the chain. Sufficient controllability criteria establish boundary control of Heisenberg and Affleck--Kennedy--Lieb--Tasaki chains~\cite{burgarth2009}. For coupled superconducting qubits, cross-resonance driving acts on one qubit at the transition frequency of another and produces tunable effective two-qubit operations~\cite{rigetti2010}. For a circulating fluxon, the passage frequency selects transitions of the exchange-coupled chain. The fluxon translation and field oscillations evolve between encounters and enter the modulation coefficients.

We consider an exchange-coupled chain whose first qubit interacts with the fluxon. Exchange transfers excitation amplitudes between the endpoint and the interior throughout each circulation. The environment of an internal qubit then contains both the fluxon position and the field modes. The distinguishability of their conditional states depends on the history of the chain. We calculate the joint response and the separate translational and radiative responses.

The observables selected by recovery or measurement determine the visible part of quantum dynamics. Krylov closure of the decoded logical observables gives a reduction for quantum codes~\cite{gordeychuk2026codes}; closure of the Fourier measurement effects gives the corresponding reduction for quantum phase estimation~\cite{gordeichuk2026qpe}. In the fluxon--chain problem, excitation probabilities and conditional overlaps assign different weights to the field branches. Repeated passages introduce a further frequency dependence through finite phase sums, whose squared modulus has the Fej\'er-kernel structure also present in phase-estimation distributions.

Perturbation theory for fluxons separates the evolution of collective parameters from field deformations~\cite{mclaughlin1978}. The modulation of spectral corrections has been derived for perturbed sine--Gordon kinks and nonlinear Klein--Gordon solitary waves~\cite{kiselev1992,kiselev2000}. Uniform elliptic-function asymptotics describe the near-separatrix profile on a long ring~\cite{kiselev2019elliptic}. Accumulation over repeated near-separatrix passages has also been studied in the perturbed Duffing problem~\cite{kiselev2012duffing}. We expand the field correction in the canonically normalized modes of the moving ring profile and quantize the translational coordinate. Both evolve between encounters.

Time ordering also affects the conditional overlap. Successive linear spin--motion interactions can generate nonlinear effective operations~\cite{sutherland2021}. In the fluxon problem, these ordered contributions coexist with the curvature of the position-dependent coupling. Both enter the conditional overlap, together with returns from the excited-mode sector. Their relative phases determine the sign of the mixed response.

We derive modulation equations for the complex amplitudes of the joint fluxon--chain state. Their coefficients are calculated from one-passage variation equations, including the vacuum response, translational dynamics, and ordered returns from excited field branches. A discrete normal form separates near-resonant amplitude evolution from the nonresonant reconstruction of the state. The resulting system determines the population and complex coherence of the internal qubit after successive encounters.

The source projections and fluxon motion determine the coefficients of the modulation system. Ordered emission and return amplitudes contribute to the internal-qubit population and coherence as the field state evolves between passages. The next response coefficient gives the diagonal and off-diagonal error contributions. A recurrence for the residual bounds their accumulation. At fixed chain exchange, velocity selects the leading resonance. Direct propagation in enlarged field and translational spaces tests the predicted populations and coherences over 81 passages. Mass and frequency variations determine the suppression intervals; spectral projections give the contributions of individual field branches.

\section{Joint fluxon--chain model}
\subsection{Ring profile and oscillatory modes}
Space and time are measured in the Josephson penetration length and inverse plasma frequency. The line energy scale is $E_0$, and $\hstar=\hbar\omega_p/E_0$. Here $\omega_p$ is the Josephson plasma angular frequency. The dimensionless circumference is $L$; the velocity $v$ is measured in units of the Josephson penetration length times $\omega_p$, with $0<v<1$. The field $\phi$ is the gauge-invariant phase difference across the junction. We write dimensionless energies as $\mathcal H=E/E_0$ and frequency Hamiltonians as $H=\mathcal H/\hstar$, so that $i\partial_t\Psi=H\Psi$. All frequencies below are measured in units of $\omega_p$. The unperturbed phase satisfies
\begin{equation}
 \phi_{tt}-\phi_{xx}+\sin\phi=0,\qquad
 \phi(x+L,t)=\phi(x,t)+2\pi .
 \label{eq:sg}
\end{equation}
A uniformly translating one-fluxon profile of Eq.~\eqref{eq:sg} is
\begin{equation}
\begin{aligned}
 \Phi_v(y)&=\pi+2\operatorname{am}\!\left(\frac{y}{k\sqrt{1-v^2}},k\right),\\
 L&=2k\sqrt{1-v^2}\,K(k),
\end{aligned}
\label{eq:profile}
\end{equation}
where $y=x-vt$, $k$ is the elliptic modulus, and $K(k)$ is the complete elliptic integral of the first kind; $\operatorname{am}$ denotes the Jacobi amplitude. Its derivative and $\cos\Phi_v$ are periodic. The large-ring limit of Eq.~\eqref{eq:profile} approaches the separatrix profile and connects to the uniform asymptotics of Ref.~\cite{kiselev2019elliptic}.

Linearization in the moving coordinate gives
\begin{equation}
\begin{aligned}
 \bigl[-(1-v^2)\partial_y^2+\cos\Phi_v
       +2iv\omega_\nu\partial_y-\omega_\nu^2\bigr]u_\nu&=0,\\
 u_\nu(y+L)&=u_\nu(y).
\end{aligned}
\label{eq:modes}
\end{equation}
The index $\nu$ labels a ring mode, $u_\nu(y)$ is its complex spatial profile, and $\omega_\nu$ is its frequency in the moving coordinate. The associated linear perturbation has time dependence $u_\nu(y)e^{-i\omega_\nu t}$. We choose positive-frequency modes with canonical normalization
\begin{equation}
 \int_{-L/2}^{L/2}\!\left[(\omega_\nu+\omega_\alpha)u_\nu^*u_\alpha
       -2iv u_\nu^*\partial_yu_\alpha\right]\dd y=\delta_{\nu\alpha}.
 \label{eq:norm}
\end{equation}
The translation mode is represented by the center coordinate. The remaining solutions of Eqs.~\eqref{eq:modes} and \eqref{eq:norm} define the discrete oscillator basis. The inertial coefficient at the chosen mean velocity is
\begin{equation}
 M_0=\frac{\dd P}{\dd v},\qquad
 P(v)=v\int_{-L/2}^{L/2}(\Phi_v')^2\dd y.
 \label{eq:mass}
\end{equation}
We use a constant effective mass $M$ for the translational fluctuation. Setting $M=M_0$ connects it to Eq.~\eqref{eq:mass}; varying $M$ at fixed $v$ defines the mass diagnostic used below.

\subsection{Source projections and Hamiltonian}
The dimensionless coupling $\kappa$ sets the interaction-energy scale in units of $E_0$ for the specified source profile. The source amplitude is included in $g$ and held fixed when $\kappa$ is varied. Let $g(x)$ be a real periodic source of zero mean and $G'=g$, with $G$ periodic and $\int_0^L G(x)\dd x=0$. In Fourier variables this convention is $G_0=0$ and $G_q=g_q/(iq)$ for $q\ne0$. It fixes the static endpoint bias: a constant shift $G\mapsto G+C$ adds $2\pi\kappa C Z_1$ to the energy. The line--qubit interaction energy is $\kappa Z_1\int G(x)\phi_x(x)\dd x$. Projection on the profile and on its oscillations gives
\begin{align}
 J(X)&=-\int_{-L/2}^{L/2}G(y+X)\Phi_v'(y)\dd y,\nonumber\\
 C_{\nu q}&=g_q\int_{-L/2}^{L/2}u_\nu^*(y)e^{iqy}\dd y,
 \qquad q=2\pi m/L,\nonumber\\
 h_\nu(X)&=-\frac{\kappa}{\sqrt{\hstar}}
                    \sum_q C_{\nu q}e^{iqX},
 \qquad g(x)=\sum_qg_qe^{iqx}.
 \label{eq:projections}
\end{align}
Here $J(X)$ is the scalar overlap with the fluxon profile, $C_{\nu q}$ is the projection of spatial Fourier harmonic $q$ onto mode $\nu$, and $h_\nu(X)$ is the corresponding oscillator coupling in frequency units. We write the center as $X=X_{\rm cl}(t)+\xi$, where $X_{\rm cl}=vt-L/2$ and $T=L/v$.
The displacement $\xi\in\mathbb R$ is a local, unwrapped coordinate about the moving reference center. The source functions are evaluated modulo $L$, while the packet is propagated on this covering coordinate. This description applies to a packet localized within a single winding branch, with negligible overlap with its translates by $L$. The Hermite representation implements this localized-packet approximation. The reference center completes one turn in time $T$; deviations from it evolve through the packet Hamiltonian. All passage values are samples of one continuous joint evolution at $t=NT$, so the packet, field modes, and chain carry their correlations into the next encounter.

The chain has number-conserving exchange and longitudinal coupling at its endpoint. The longitudinal interaction changes the endpoint excitation energy. Its nonzero commutator with exchange allows transitions between chain eigenstates at fixed total excitation number, changing the internal-qubit population. We use the rotating frame of the common transition frequency of the resonant qubits. The common excitation-energy term is removed in this frame; the exchange and $Z_1$ coupling are unchanged. In frequency units its Hamiltonian is
\begin{equation}
 H_q=\sum_{j=1}^{n_q-1}\mathcal J_j
       (\sigma_j^+\sigma_{j+1}^-+\sigma_{j+1}^+\sigma_j^-),
 \qquad Z_1=\sigma_1^z.
 \label{eq:chain}
\end{equation}
We denote the number of retained field modes by $K$ and the translational mass by $M$. The numerical example uses three resonant qubits, $\mathcal J_1=\mathcal J_2=\mathcal J$, and the vacuum plus one-excitation sector. In the basis $|0\rangle,|1\rangle,|2\rangle,|3\rangle$, where $|0\rangle$ is the chain vacuum and $|j\rangle$ has one excitation on qubit $j$,
\begin{equation}
 H_q=\mathcal J
 \begin{pmatrix}0&0&0&0\\0&0&1&0\\0&1&0&1\\0&0&1&0\end{pmatrix},
 \qquad Z_1=\operatorname{diag}(1,-1,1,1).
 \label{eq:matrices}
\end{equation}
Here $n_q$ is the number of qubits, $\mathcal J_j$ is the exchange frequency on bond $(j,j+1)$, and $\sigma_j^+$ and $\sigma_j^-$ create and remove a site excitation. Our convention is $\sigma_j^z|0_j\rangle=|0_j\rangle$ and $\sigma_j^z|1_j\rangle=-|1_j\rangle$. The matrices in Eq.~\eqref{eq:matrices} implement this invariant sector of Eq.~\eqref{eq:chain}.

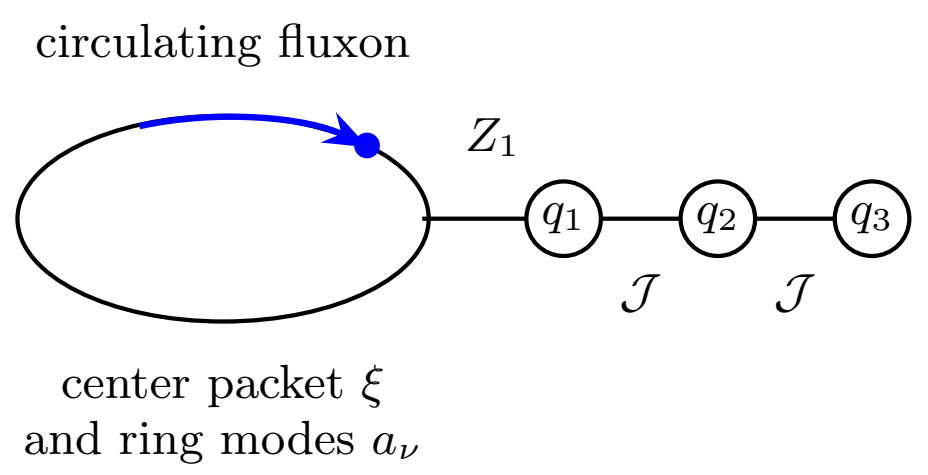
\begin{figure}
\centering
\begin{tikzpicture}[scale=.85,>=Stealth]
 \draw[thick] (0,0) ellipse (1.6 and .8);
 \draw[very thick,->,blue] (-.65,.72) arc[start angle=114,end angle=45,x radius=1.6,y radius=.8];
 \fill[blue] (1.12,.57) circle (.10);
 \node[above] at (0,1.05) {circulating fluxon};
 \draw[thick] (1.55,0)--(2.35,0);
 \foreach \x/\j in {2.65/1,3.85/2,5.05/3}{
  \draw[fill=white,thick] (\x,0) circle (.29);
  \node at (\x,0) {$q_{\j}$};}
 \draw[thick] (2.94,0)--(3.56,0);
 \draw[thick] (4.14,0)--(4.76,0);
 \node[below] at (3.25,-.3) {$\mathcal J$};
 \node[below] at (4.45,-.3) {$\mathcal J$};
 \node[below,align=center] at (0,-1.0) {center packet $\xi$\\and ring modes $a_\nu$};
 \node[above] at (2.1,.35) {$Z_1$};
\end{tikzpicture}
\caption{The fluxon position and the ring modes couple to the endpoint operator $Z_1$. Exchange acts throughout the evolution. The conditional overlap of the internal qubit $q_2$ is evaluated after each period.}
\label{fig:scheme}
\end{figure}

The oscillator operators satisfy $[a_\nu,a_\alpha^\dagger]=\delta_{\nu\alpha}$; $a_\nu^\dagger a_\nu$ counts quanta of mode $\nu$. The joint state belongs to the tensor product of the packet space $L^2(\mathbb R_\xi)$, the oscillator Fock space, and the retained chain sector. Each factor-specific operator is extended by the identity on the other factors. A dagger denotes the Hermitian adjoint, and a star denotes complex conjugation. The joint frequency Hamiltonian is
\begin{align}
 H_0(t)&=-\frac{\hstar}{2M}\partial_\xi^2+H_q
       -\frac{\kappa}{\hstar}J(X_{\rm cl}+\xi)Z_1,\nonumber\\
 H(t)&=H_0(t)+\sum_\nu\omega_\nu a_\nu^\dagger a_\nu\nonumber\\
 &\quad+\lambda\sum_\nu\left[h_\nu(X_{\rm cl}+\xi)a_\nu^\dagger
                   +h_\nu^*(X_{\rm cl}+\xi)a_\nu\right]Z_1.
 \label{eq:hamiltonian}
\end{align}
The parameter $\lambda$ counts radiative vertices and is set to one in numerical comparisons. Equation~\eqref{eq:hamiltonian} retains the full position dependence of $J$ and $h_\nu$ within the harmonic-mode, constant-mass collective-coordinate model. Its coefficients are periodic in time. Appendix~\ref{app:field-reduction} derives the collective kinetic metric from the field action and gives the geometric interaction terms. Figure~\ref{fig:scheme} specifies the coupling arrangement.

Equation~\eqref{eq:hamiltonian} is equivalent to the following coupled system for four-component Fock amplitudes $\Psi_{\boldsymbol n}(\xi,t)$:
\begin{align}
 i\partial_t\Psi_{\boldsymbol n}
 &=\left[H_0(t)+\sum_\nu n_\nu\omega_\nu\right]\Psi_{\boldsymbol n}\nonumber\\
 &\quad+\lambda Z_1\sum_\nu\bigl[
 h_\nu\sqrt{n_\nu}\,\Psi_{\boldsymbol n-\boldsymbol e_\nu}
 +h_\nu^*\sqrt{n_\nu+1}\,\Psi_{\boldsymbol n+\boldsymbol e_\nu}\bigr].
 \label{eq:system}
\end{align}
The multi-index $\boldsymbol n=(n_1,\ldots,n_K)$ lists oscillator occupations, and $\boldsymbol e_\nu$ increments occupation $\nu$ by one. Amplitudes with a negative occupation are zero. The initial condition for Eq.~\eqref{eq:system} is
\begin{equation}
 \Psi_{\boldsymbol0}(\xi,0)=\varphi(\xi)c_{\rm in},\quad
 \varphi=\frac{e^{-\xi^2/(4\sigma_X^2)}}{(2\pi\sigma_X^2)^{1/4}},\quad
 c_{\rm in}=\begin{pmatrix}1/\sqrt2\\1/\sqrt6\\1/\sqrt6\\1/\sqrt6\end{pmatrix},
 \label{eq:initial}
\end{equation}
Here $\sigma_X$ is the standard deviation of the initial center position; the real Gaussian has zero mean momentum and momentum standard deviation $\hstar/(2\sigma_X)$. All excited-mode amplitudes are initially zero. The chain preparation is $(|0\rangle+|W_3\rangle)/\sqrt2$, where $|W_3\rangle=(|1\rangle+|2\rangle+|3\rangle)/\sqrt3$. This state has nonzero vacuum--excitation coherence at each site.

\subsection{Conditional coherence and the mixed contribution}
Tracing the center and the oscillators gives the chain density matrix
\begin{equation}
\begin{aligned}
 \rho_{jk}&=\sum_{\boldsymbol n}\int
       \Psi_{\boldsymbol n,j}(\xi)\Psi_{\boldsymbol n,k}^*(\xi)\dd\xi,\\
 \eta_j&=\frac{|\rho_{j0}|}{\sqrt{\rho_{jj}\rho_{00}}},\qquad
 \Gamma_j=-\log\eta_j.
\end{aligned}
\label{eq:observable}
\end{equation}
We use $\eta_j$ and $\Gamma_j$ when $\rho_{jj}\rho_{00}>0$, with finite $\Gamma_j$ for nonzero overlap. The amplitudes conditional on chain states $j$ and $0$ are environmental vectors; hence $0\leq\eta_j\leq1$. In the invariant sector, the off-diagonal element of the local density matrix of qubit $j$ is $\rho_{j0}$. Its magnitude is $\eta_j\sqrt{\rho_{jj}\rho_{00}}$. Equation~\eqref{eq:observable} separates conditional environmental distinguishability from the population factor and coherent phase evolution. The normalized fluxon--mode states conditional on $|j\rangle$ and $|0\rangle$ have overlap modulus $\eta_j$. Exchange changes both their overlap and the chain populations.

We compare three evolutions with the same initial chain state: $X$, the translational packet with $\lambda=0$; $R$, radiative modes along the prescribed center trajectory $\xi=0$; and $XR$, their joint evolution. The signed mixed response is
\begin{equation}
 \Gamma_{{\rm mix},j}=\Gamma_{XR,j}-\Gamma_{X,j}-\Gamma_{R,j}.
 \label{eq:mixed}
\end{equation}
Independent multiplication of the two conditional overlaps gives zero in Eq.~\eqref{eq:mixed}. Positive and negative values describe enhanced and suppressed conditional-coherence loss relative to that product. We report $j=2$ below.

\section{Resummed modulation at fixed exchange coupling}
\label{sec:resummed-modulation}
\subsection{Resonance selected by the fluxon velocity}
The modulation calculation uses the nonlinear extension of the collective model in Eq.~\eqref{eq:hamiltonian}. We retain the cubic and quartic field energies, the source through quadratic order in transformed field fluctuations, and the canonical momentum-dependent vertices. The translational coordinate in the source profile remains unexpanded. The complete energy is given in Eq.~\eqref{eq:full-field-four}. Frequency and momentum counterterms then set the source-free excitation gaps and the translational curvature to prescribed values, as in Eq.~\eqref{eq:matched-hamiltonian}. These choices define the Hamiltonian used for both the modulation construction and its direct numerical reference.

The initial oscillator state for this nonlinear model is its matched source-free ground state at zero momentum. It is multiplied by the same Gaussian packet and chain vector as in Eq.~\eqref{eq:initial}. The harmonic model of Sec.~\ref{sec:harmonic-response} uses the harmonic vacuum and expresses the correction in terms of mode emission and return amplitudes.

We use the following expansion parameters. The physical coupling is $\kappa$. Field degree counts powers of canonical fluctuations in the Hamiltonian. The parameters $\epsilon_{\rm v}$ and $\epsilon_{\rm t}$ below count vacuum and translational response insertions within the ground field band; $\zeta$ counts transitions between that band and its complement. These three response parameters are set to one after their coefficients have been calculated, with $\kappa$ fixed. The diagonal evolution contains the periodic source at its physical strength. The order of each approximation below refers to the specified response parameter.

At fixed exchange coupling, we select the circulation period by the chain resonance condition. The frequency Hamiltonian is $H=\mathcal H^{\rm match}/\hstar$, with the field and source terms of Eq.~\eqref{eq:full-field-four} and the matching prescription of Appendix~\ref{app:physical-matching}. In this calculation the target mode gaps are the natural frequencies of the moving ring profile at the selected velocity. The source geometry is specified in Eq.~\eqref{eq:source}. Its Fourier coefficients are
\[
 g_q=\frac{2A_g}{L}
 \frac{\cos(qa)+r\cos(qb)-(1+r)\cos(qc)}
 {\cosh(\pi wq/2)}.
\]
Here $q=2\pi m/L$, $m\in\mathbb Z$. The parameters $a,b,c$ give the lobe positions, $w$ their width, $r$ the relative weight, and $A_g$ the source amplitude; their numerical values are listed with Eq.~\eqref{eq:source}. The nonzero chain energies of Eq.~\eqref{eq:matrices} are $E_\pm=\pm\sqrt2\mathcal J$, and $\langle -|Z_1|+\rangle=-1/2$. Equating their separation to the first circulation harmonic gives
\begin{equation}
 \Omega=\frac{2\pi v}{L}=2\sqrt2\mathcal J,
 \qquad v_{\rm res}=\frac{\sqrt2\mathcal J L}{\pi},
 \qquad T_{\rm res}=\frac{\pi}{\sqrt2\mathcal J}.
 \label{eq:velocity-resonance}
\end{equation}
Equation~\eqref{eq:velocity-resonance} specifies the leading resonant pair; the coefficient calculations below retain all temporal harmonics with $|m|\leq18$. The parameters are
\[
\begin{gathered}
 \mathcal J=0.018,\quad L=12,\quad \kappa=10^{-5},\\
 v=0.0972341621,\quad T=123.413415,\\
 \sigma_X=0.2,\quad \hstar=5.96262690\times10^{-4},\\
 M=126.605671.
\end{gathered}
\]
Here $M=M_0/\mu$, with $\mu=0.0641125286$, is a prescribed effective mass in the curvature family. The ring profile determines $M_0$; the chosen $\mu$ fixes the kinetic coefficient of this numerical model independently. Matching preserves that chosen curvature as the retained field basis changes. The same value of $\mu$ is used in the passage-27 diagnostic of Appendix~\ref{app:mass-response}, while the present resonance test uses the velocity and coupling listed above. The first three natural mode frequencies are $1.01370675$, $1.08522575$, and $1.25475807$. The initial state is the product of the Gaussian packet, the matched mode ground state at zero momentum, and the chain state in Eq.~\eqref{eq:initial}.

The large-ring leading profile gives
\[
 \int_{-\infty}^{\infty}\Phi'_v(x)e^{iqx}\dd x
 =\frac{2\pi}{\cosh(\pi q\sqrt{1-v^2}/2)}.
\]
Together with Eq.~\eqref{eq:projections}, this yields the resonant matrix element $-i\kappa\mathfrak g(v)e^{iq\xi}$, where $q=2\pi/L$ and
\[
 \mathfrak g(v)=\frac{\pi g_q}
 {\hstar q\cosh(\pi q\sqrt{1-v^2}/2)}.
\]
Quadrature on the ring gives $\mathfrak g_{\rm cl}=370.9014876$, compared with $370.8967370$ from the isolated-kink expression. At $\xi=0$, the leading rotating-frame amplitudes satisfy
\begin{equation}
 i\frac{\dd}{\dd t}\begin{pmatrix}b_-\\b_+\end{pmatrix}
 =\kappa\mathfrak g_{\rm cl}
 \begin{pmatrix}0&-i\\i&0\end{pmatrix}
 \begin{pmatrix}b_-\\b_+\end{pmatrix}.
 \label{eq:resonant-pair-modulation}
\end{equation}
The amplitudes $b_-$ and $b_+$ multiply the chain eigenstates with energies $E_-$ and $E_+$ after their free phases have been removed. The phase accumulated in Eq.~\eqref{eq:resonant-pair-modulation} is $\vartheta_N=\kappa\mathfrak g_{\rm cl}NT$. For the specified chain preparation this gives the leading population
\[
 P_2(N)=\frac{[\cos\vartheta_N+\sqrt2\sin\vartheta_N]^2}{6}.
\]
It identifies the first-order slow scale $\kappa t$ and the amplitude exchange to which the field corrections are added. The corresponding uncoupled-chain control is known exactly: $c^{(0)}(t)=e^{-iH_qt}c_{\rm in}$. At the resonance in Eq.~\eqref{eq:velocity-resonance}, the excited chain eigenstates acquire the same factor $(-1)^N$ at $t=NT$, so
\[
 P_2^{(0)}(N)=\frac16,\qquad
 c_2^{(0)}(N)=\frac{(-1)^N}{\sqrt{12}}.
\]
These stroboscopic values give the reference for the total fluxon-induced change. The leading two-state formula displays the resonance analytically; the modulation construction below retains the remaining source harmonics, packet dynamics, and field corrections.

\subsection{Vacuum and translational response coefficients}
Let $\Pi$ denote the retained source-free ground field band, including its translational states. This band is the low-energy subspace continuously connected to the mode ground state with translational excitations; momentum-dependent couplings dress these product states. Its coordinate representation is fixed by the unitary polar factor of its overlap with the product of the matched mode vacuum and the packet coordinate basis. Write the projected Hamiltonian in this representation as
\begin{equation}
 H_{\rm ad}(t)=H_{\rm cl}(t)+\delta H_{\rm v}(t)
                                  +\delta H_{\rm t}(t).
 \label{eq:adiabatic-response-split}
\end{equation}
Here $H_{\rm cl}$ contains $H_q$ and the classical periodic source at the frozen coordinate $\xi$. The vacuum correction $\delta H_{\rm v}$ is the change of the source after taking its expectation in the matched mode ground state. The translational correction $\delta H_{\rm t}$ contains the band kinetic operator and the remaining momentum-dependent source terms. Its leading kinetic term is $p^2/(2M\hstar)$, with $[\xi,p]=i\hstar$. This construction includes the cubic and quartic field terms through the matched eigenstates and projected vertices.

The leading source coefficient changes to $\mathfrak g_{\rm v}=370.8830694$, so $\delta\mathfrak g=-0.01841825$. Consequently the leading vacuum phase correction is $\delta\vartheta_N=\kappa\delta\mathfrak g NT$. Its Taylor expansion in a population or coherence contains terms proportional to $N$ and $N^2$. We compute the corresponding operator corrections by inserting independent response parameters $\epsilon_{\rm v},\epsilon_{\rm t}$ in Eq.~\eqref{eq:adiabatic-response-split}. The physical coupling $\kappa$ retains its specified value in each coefficient.

Expand the one-period evolution as
\[
 F(\epsilon_{\rm v},\epsilon_{\rm t})
 =\sum_{a,b\geq0}\epsilon_{\rm v}^a\epsilon_{\rm t}^b F_{ab}.
\]
The coefficients through $a+b=2$ follow from
\begin{equation}
\begin{split}
 i\dot U_{ab}={}&H_{\rm cl}U_{ab}
 +\delta H_{\rm v}U_{a-1,b}+\delta H_{\rm t}U_{a,b-1},\\
 U_{00}(0)={}&I,\qquad U_{ab}(0)=0\quad(a+b>0),\\
 F_{ab}={}&U_{ab}(T).
\end{split}
\label{eq:bivariate-period-variation}
\end{equation}
Negative indices in Eq.~\eqref{eq:bivariate-period-variation} denote zero. This coefficient convention includes the Taylor factorials. $F_{11}$ contains both orders of a vacuum and a translational insertion.

For one response parameter, write $F=F_{00}+\epsilon F_1+\epsilon^2F_2+\cdots$. The linear term in $F^N$ is
\[
 \sum_{n=0}^{N-1}F_{00}^{N-1-n}F_1F_{00}^{n}.
\]
If $F_{00}|\alpha\rangle=e^{-i\theta_\alpha}|\alpha\rangle$, its matrix element contains
\begin{equation}
\begin{aligned}
 D_N(\Delta)&=\sum_{n=0}^{N-1}e^{in\Delta}\\
 &=e^{i(N-1)\Delta/2}\frac{\sin(N\Delta/2)}{\sin(\Delta/2)},\\
 \Delta&=\theta_\alpha-\theta_\beta.
\end{aligned}
 \label{eq:response-passage-sum}
\end{equation}
The continuous value of Eq.~\eqref{eq:response-passage-sum} at $\Delta=2\pi l$ is $N$. Two resonant insertions produce $N(N-1)/2$. These factors describe the secular terms that are resummed in the generator.

Choose a quasiphase branch for $F_{00}=W_0e^{-i\Theta}W_0^\dagger$ on which the required exponential divided differences are nonzero. The columns of $W_0$ are orthonormal eigenvectors and $\Theta=\operatorname{diag}(\theta_\alpha)$ contains their quasiphases in radians per passage. We use divided differences $f[x,y]=(f(x)-f(y))/(x-y)$ and $f[x,y,z]=(f[x,y]-f[y,z])/(x-z)$, with derivative limits at coincident arguments. Put $f(x)=e^{-ix}$ and $\widehat F_{ab}=W_0^\dagger F_{ab}W_0$. The first generator coefficients obey
\[
 (G_x)_{\alpha\beta}
 =\frac{(\widehat F_x)_{\alpha\beta}}
 {f[\theta_\alpha,\theta_\beta]},\qquad x\in\{{\rm v},{\rm t}\},
\]
where $F_{\rm v}=F_{10}$ and $F_{\rm t}=F_{01}$. The pure quadratic coefficient is
\[
 (G_{xx})_{\alpha\beta}
 =\frac{(\widehat F_{xx})_{\alpha\beta}
 -\sum_\gamma f[\theta_\alpha,\theta_\gamma,\theta_\beta]
 (G_x)_{\alpha\gamma}(G_x)_{\gamma\beta}}
 {f[\theta_\alpha,\theta_\beta]}.
\]
Here $F_{\rm vv}=F_{20}$ and $F_{\rm tt}=F_{02}$. For the mixed coefficient, replace the product in the sum by
$(G_{\rm v})_{\alpha\gamma}(G_{\rm t})_{\gamma\beta}
 +(G_{\rm t})_{\alpha\gamma}(G_{\rm v})_{\gamma\beta}$ and use $F_{11}$. Confluent divided differences give the limits at repeated quasiphases. The resummed discrete evolution in the $W_0$ basis is $a_{n+1}=e^{-iG_{\rm ad}}a_n$, where $n\in\mathbb N_0$ counts completed passages. For the chosen generator define its continuous interpolation by $a(s)=e^{-isG_{\rm ad}}a_0$, with $s\in[0,\infty)$ and $a_0=W_0^\dagger\Psi_{\rm ad,in}$. Differentiation with respect to the real variable $s$ gives
\begin{equation}
 i\frac{\dd a}{\dd s}=G_{\rm ad}a,\qquad
 G_{\rm ad}=\Theta+G_{\rm v}+G_{\rm t}
                 +G_{\rm vv}+G_{\rm vt}+G_{\rm tt}.
 \label{eq:adiabatic-resummed-generator}
\end{equation}
The vector $a(s)$ contains ground-band and chain amplitudes in the $W_0$ basis, and $G_{\rm ad}$ is dimensionless, in radians per passage. The identity $a(s)|_{s=n}=a_n$ matches the interpolation to the discrete sequence at the sampling times $t=nT$. After $N\in\mathbb N_0$ passages the reconstructed state is $W_0e^{-iNG_{\rm ad}}W_0^\dagger\Psi_{\rm ad,in}$. Evolution within a period is determined by the time-dependent Hamiltonian. The prepared band state $\Psi_{\rm ad,in}$ is the normalized projection of the incident packet. Equation~\eqref{eq:adiabatic-resummed-generator} preserves the noncommuting mixed response and sums its repeated insertions.

\subsection{Excited branches and ordered return amplitudes}
We next retain the excited field sector $Q=I-\Pi$ and its state between passages. Both projectors act as the identity on the chain; $I$ is the identity on the retained joint space. In this section $Q$ denotes the complementary field projector. For the full matched frequency Hamiltonian define
\begin{equation}
\begin{gathered}
 H_{\rm d}=\Pi H\Pi+QHQ,\qquad V=\Pi HQ+QH\Pi,\\
 H_\zeta=H_{\rm d}+\zeta V.
\end{gathered}
 \label{eq:interband-split}
\end{equation}
The auxiliary parameter $\zeta$ counts transitions between the two field sectors and is set to one. The diagonal blocks in Eq.~\eqref{eq:interband-split} retain the chain exchange, translation, and all source harmonics. The ground block equals the adiabatic Hamiltonian up to the coordinate transformation above. We integrate both diagonal blocks directly for the following comparison; the previously resummed adiabatic solution agrees with its direct control to better than $10^{-10}$ in the tested observables.

Let $U_{\rm d}(t)$ be the diagonal-block evolution and put $V_I(t)=U_{\rm d}^\dagger(t)V(t)U_{\rm d}(t)$. Expand the full period operator as $F_{\rm d}+\zeta F^{(1)}+\zeta^2F^{(2)}+\cdots$. The coefficients are obtained on one period from
\begin{equation}
\begin{aligned}
 i\dot U^{(0)}&=H_{\rm d}U^{(0)},& U^{(0)}(0)&=I,\\
 i\dot U^{(1)}&=H_{\rm d}U^{(1)}+VU^{(0)},& U^{(1)}(0)&=0,\\
 i\dot U^{(2)}&=H_{\rm d}U^{(2)}+VU^{(1)},& U^{(2)}(0)&=0.
\end{aligned}
\label{eq:interband-variation}
\end{equation}
Here $F_{\rm d}=U^{(0)}(T)$ and $F^{(j)}=U^{(j)}(T)$. Define $A_j=F_{\rm d}^\dagger F^{(j)}$. Unitarity of the coefficient expansion gives $A_1+A_1^\dagger=0$ and $A_2+A_2^\dagger+A_1^\dagger A_1=0$. Therefore
\begin{equation}
\begin{aligned}
 K_1&=iA_1=\int_0^T V_I(t)\dd t,\\
 K_2&=i\left(A_2-\tfrac12A_1^2\right)\\
 &=-\frac{i}{2}\int_0^T\dd t\int_0^t\dd s\,[V_I(t),V_I(s)].
\end{aligned}
 \label{eq:interband-kick}
\end{equation}
are Hermitian. We compute them by integrating Eq.~\eqref{eq:interband-variation}; the commutator in Eq.~\eqref{eq:interband-kick} accounts for time ordering.

The operator $K_1$ is off diagonal in $\Pi+Q$, and $K_2$ is block diagonal. The ground-band correction is
\begin{equation}
\begin{split}
 \Pi K_2\Pi=-\frac{i}{2}\int_0^T\dd t\int_0^t\dd s\,
 \bigl[&\Pi V_I(t)Q V_I(s)\Pi\\
       &-\Pi V_I(s)Q V_I(t)\Pi\bigr].
\end{split}
\label{eq:mode-return-generator}
\end{equation}
Resolution of $Q$ into field states expresses Eq.~\eqref{eq:mode-return-generator} as a sum over intermediate branches. Their frequencies, packet dependence, and the chain dynamics enter the two ordered factors. Exponentiation of $K_1$ already produces a quadratic return contribution; $K_2$ adjusts it to the actual ordered return amplitude. This distinction determines the comparison below.

\subsection{Passage modulation and reconstruction of observables}
The second-order unitary passage approximation is
\begin{equation}
 F_{\rm eff}=F_{\rm d}\exp[-i(K_1+K_2)],\qquad
 \Psi_N=F_{\rm eff}^{N}\Psi_{\rm in}.
 \label{eq:full-resummed-passage}
\end{equation}
The ordered product in Eq.~\eqref{eq:full-resummed-passage} retains the commutators with intraband evolution. All coefficients are determined over $0\leq t\leq T$; the state in both sectors is then carried through the sequence. In the physical calculation $V$ is proportional to $\kappa$, while $U_{\rm d}$ already includes the response to that coupling. The expansion in $\zeta$ organizes interband transitions around this evolving reference.

The passage map acts at integer indices: $\Psi_{n+1}=F_{\rm eff}\Psi_n$, $n\in\mathbb N_0$. In the retained finite-dimensional space, diagonalize it as $F_{\rm eff}=W\operatorname{diag}(e^{-i\vartheta_\alpha})W^\dagger$ and fix real representatives $\vartheta_\alpha$ of the eigenphases. Put $\Theta_{\rm eff}=\operatorname{diag}(\vartheta_\alpha)$. The sequence $a_n=W^\dagger\Psi_n$ then satisfies $a_{n+1}=e^{-i\Theta_{\rm eff}}a_n$. Define the differentiable interpolation $a(s)=e^{-is\Theta_{\rm eff}}W^\dagger\Psi_{\rm in}$ for real $s\geq0$. It satisfies
\begin{equation}
\begin{gathered}
 i\frac{\dd a(s)}{\dd s}=\Theta_{\rm eff}a(s),\qquad s\geq0,\\
 \Psi_n=Wa(s)\big|_{s=n},\qquad n\in\mathbb N_0,\\
 G_{\rm eff}=W\Theta_{\rm eff}W^\dagger.
\end{gathered}
 \label{eq:full-stroboscopic-generator}
\end{equation}
Equation~\eqref{eq:full-stroboscopic-generator} is an exact interpolation of the approximate passage map in Eq.~\eqref{eq:full-resummed-passage}: $We^{-in\Theta_{\rm eff}}W^\dagger=F_{\rm eff}^{n}$ for every integer $n\geq0$. The choice of real eigenphases fixes a Hermitian logarithm. Replacing $\vartheta_\alpha$ by $\vartheta_\alpha+2\pi k_\alpha$, $k_\alpha\in\mathbb Z$, changes the interpolation at noninteger $s$ and preserves all sampled states. The calculations below compare these integer-passage states. The slow resonant amplitudes are separated by the discrete normal form in Sec.~\ref{sec:resonant-normal-form}. Mode memory enters through the eigenphases and the intersector amplitudes. Let $F$ denote the exact one-period propagator of the retained Hamiltonian. For unitary exact and approximate passage maps, the telescoping identity gives
\[
 \|F_{\rm eff}^N-F^N\|\leq N\|F_{\rm eff}-F\|.
\]
Here and below $\|\cdot\|$ is the Hilbert-space norm for vectors and the induced spectral norm for operators; $\|\cdot\|_F$ denotes the Frobenius matrix norm. The inequality relates accumulated error to the local map error, with observable accuracy assessed on the specified incident state.

Let $b_{\alpha j}(N)$ be amplitudes in an orthonormal field basis and the chain basis of Eq.~\eqref{eq:matrices}. We reconstruct
\begin{equation}
\begin{gathered}
 \rho_{jk}(N)=\sum_\alpha b_{\alpha j}(N)b_{\alpha k}^*(N),\\
 P_2=\rho_{22},\qquad c_2=\rho_{02}.
\end{gathered}
 \label{eq:modulation-observables}
\end{equation}
$c_2$ is the complex conjugate of the coherence orientation used in Eq.~\eqref{eq:observable}. Both field sectors contribute to Eq.~\eqref{eq:modulation-observables}. The local qubit density matrix is $\bigl(\begin{smallmatrix}1-P_2&c_2\\c_2^*&P_2\end{smallmatrix}\bigr)$. Comparing complex $c_2$ tests its phase and modulus simultaneously.

\subsection{A unified approximate passage operator}
\label{sec:unified-passage}
We combine the two response constructions in the same field representation. Let $\mathcal S$ be the unitary map from the source-free ground-band eigenbasis and chain energy basis to the packet-coordinate and site basis used for $G_{\rm ad}$. If $e_0$ is the source-free field ground energy in frequency units, define
\begin{equation}
\begin{aligned}
 B_{\rm ad}&=e^{-ie_0T}\mathcal S^\dagger W_0
                   e^{-iG_{\rm ad}}W_0^\dagger\mathcal S,\\
 \widetilde F_{\rm d}&=B_{\rm ad}\oplus F_Q,\\
 \widetilde F&=\widetilde F_{\rm d}e^{-i(K_1+K_2)}.
\end{aligned}
\label{eq:unified-passage}
\end{equation}
The symbol $\oplus$ denotes the direct sum on the ground and excited field subspaces. In the corresponding block representation,
\[
\begin{gathered}
 \widetilde F_{\rm d}=
 \begin{pmatrix}B_{\rm ad}&0\\0&F_Q\end{pmatrix},\qquad
 \Psi=\begin{pmatrix}\Psi_{\rm g}\\\Psi_Q\end{pmatrix},\\
 \widetilde F_{\rm d}\Psi=
 \begin{pmatrix}B_{\rm ad}\Psi_{\rm g}\\F_Q\Psi_Q\end{pmatrix}.
\end{gathered}
\]
Here $\Psi_{\rm g}=\Pi\Psi$ and $\Psi_Q=Q\Psi$ are the components in the two subspaces. Each component includes the translational and chain degrees of freedom. The operator $B_{\rm ad}$ propagates the ground component and $F_Q$ the excited component. Transitions between them enter through $e^{-i(K_1+K_2)}$ in Eq.~\eqref{eq:unified-passage}. The off-diagonal operator $K_1$ transfers amplitudes between the two subspaces; the block-diagonal operator $K_2$ supplies the second-order time-ordering correction to emission and return. Both are computed from Eq.~\eqref{eq:interband-kick}. Equation~\eqref{eq:unified-passage} uses the quadratic vacuum--translation generator in the occupied ground band and retains excited-sector propagation between visits. The scalar phase $e^{-ie_0T}$ aligns the energy convention in the two constructions. Using direct ground-band evolution in this transformation reproduces the previously integrated diagonal block to $2.43\times10^{-12}$ in Frobenius norm. All joint comparisons in this subsection use the same 12-state translational representation and field rank 128.

The interband coefficients are evaluated about the physical diagonal-block evolution. This defines a composite approximation to the physical passage map: the ground-band replacement and the interband truncation have separately computable local defects. Their product in Eq.~\eqref{eq:unified-passage} specifies the mixed insertions throughout the sequence.

\subsection{Finite-time resonant normal form}
\label{sec:resonant-normal-form}
The resonant pair of Eq.~\eqref{eq:resonant-pair-modulation} is already retained in the diagonal-block evolution. We now separate the slower interband and return corrections. Diagonalize $\widetilde F_{\rm d}=W_{\rm d}DW_{\rm d}^\dagger$, where $D_{\alpha\alpha}=e^{-i\theta_\alpha}$, and put $k_j=W_{\rm d}^\dagger K_jW_{\rm d}$. Define the phase difference $\Delta_{\alpha\beta}$ modulo $2\pi$ in $(-\pi,\pi]$. For a sequence of length $N_*=81$, the resonant mask is
\[
 \chi_{\alpha\beta}=
 \begin{cases}1,&|\Delta_{\alpha\beta}|\leq\pi/N_*,\\
 0,&|\Delta_{\alpha\beta}|>\pi/N_*.
 \end{cases}
\]
These retained pairs have a relative phase excursion of at most $\pi$ over the observation window. The mask specifies the retained entries of the amplitude system. Its width is tested below.

For the eliminated entries, solve the first discrete homological equation:
\begin{equation}
 (S_1)_{\alpha\beta}
 =\frac{(1-\chi_{\alpha\beta})(k_1)_{\alpha\beta}}
 {e^{i\Delta_{\alpha\beta}}-1},\qquad
 R_1=\chi\circ k_1.
 \label{eq:discrete-homological}
\end{equation}
The symbol $\circ$ denotes entrywise multiplication; $S_1$ is zero on the retained entries. Hermiticity follows by interchanging $\alpha$ and $\beta$. Put $S_1^+=D^\dagger S_1D$. Expanding
$e^{i\zeta S_1}D e^{-i(\zeta k_1+\zeta^2k_2)}e^{-i\zeta S_1}$
through second order gives the corrected second-order coefficient
\begin{equation}
 L_2=k_2+\frac{i}{2}
 \bigl([S_1^+,k_1]+[S_1^+,S_1]-[k_1,S_1]\bigr).
 \label{eq:normal-second-coefficient}
\end{equation}
Apply Eq.~\eqref{eq:discrete-homological} to $L_2$ to obtain $S_2$, and set $R_2=\chi\circ L_2$. The near-identity transformation and the resonant period map are
\begin{equation}
 \mathcal T=e^{-iS_1}e^{-iS_2},\qquad
 F_{\rm R}=W_{\rm d}\mathcal T D e^{-i(R_1+R_2)}
                  \mathcal T^\dagger W_{\rm d}^\dagger.
 \label{eq:resonant-period-map}
\end{equation}
Equations~\eqref{eq:normal-second-coefficient} and \eqref{eq:resonant-period-map} retain the second-order effects of eliminated first-order transitions.

In the interaction representation the discrete modulation equations are
\begin{equation}
\begin{aligned}
 a_{n+1}&=\exp[-i\mathcal R(n)]a_n,\\
 \mathcal R(n)&=D^{-n}(R_1+R_2)D^n,\\
 \Psi_n&=W_{\rm d}\mathcal T D^n a_n,\qquad
 a_0=\mathcal T^\dagger W_{\rm d}^\dagger\Psi_{\rm in}.
\end{aligned}
\label{eq:discrete-resonant-modulation}
\end{equation}
Every matrix element in $\mathcal R(n)$ carries the slow factor $e^{in\Delta_{\alpha\beta}}$ of a retained pair. Equation~\eqref{eq:discrete-resonant-modulation} evolves the resonant amplitudes and restores the nonresonant dressing through $\mathcal T$. The variable $n$ counts complete passages; the slowly varying amplitudes evolve after the rapid diagonal phases $D^n$ have been extracted. The finite-time window retains all pairs whose phase differences accumulate slowly on the chosen observation interval.

At the chosen width, the near-identity transformations are small: $\|S_1\|_F=3.87\times10^{-4}$ and $\|S_2\|_F=1.15\times10^{-6}$. The resonant system reproduces the unified-map observables to $3.2\times10^{-13}$ over 81 passages. Halving or doubling the phase window preserves this agreement, and scaling the coefficients verifies the cubic remainder of the transformation. Appendix~\ref{app:normal-form-checks} gives these checks.

\subsection{Coefficient-derived residual and its accumulation}
\label{sec:unified-leading-residual}
For the next ground-band coefficient introduce one common parameter $\epsilon$ multiplying $\delta H_{\rm v}+\delta H_{\rm t}$. Extend Eq.~\eqref{eq:bivariate-period-variation} to third order along this direction and denote the resulting coefficient by $F_{{\rm g},3}$. In the coordinate representation let
\[
 B_2(\epsilon)=W_0\exp[-i(\Theta+\epsilon G_1+\epsilon^2G_2)]W_0^\dagger,
\]
where $G_1=G_{\rm v}+G_{\rm t}$ and $G_2=G_{\rm vv}+G_{\rm vt}+G_{\rm tt}$. The leading ground-band map defect is
\begin{equation}
 R_{{\rm g},3}=F_{{\rm g},3}-[\epsilon^3]B_2(\epsilon).
 \label{eq:ground-cubic-defect}
\end{equation}
The notation $[\epsilon^3]$ extracts the coefficient of $\epsilon^3$ in the Taylor series. The coefficient extraction in Eq.~\eqref{eq:ground-cubic-defect} is evaluated by a triangular block exponential. In the 12-state translational representation, $\|R_{{\rm g},3}\|_F=4.02423\times10^{-7}$; the fourth-order coefficient of the same defect has norm $1.41269\times10^{-8}$. These operator norms include weakly populated packet directions. The observable contribution is computed with the prepared state.

For the interband contribution extend Eq.~\eqref{eq:interband-variation} by
\[
 i\dot U^{(3)}=H_{\rm d}U^{(3)}+VU^{(2)},\qquad U^{(3)}(0)=0,
\]
and define $A_3=F_{\rm d}^\dagger U^{(3)}(T)$. The cubic Hermitian coefficient is
\begin{equation}
 K_3=i\left[A_3-\tfrac12(A_1A_2+A_2A_1)+\tfrac13A_1^3\right].
 \label{eq:interband-cubic-coefficient}
\end{equation}
The sector parity makes $K_3$ off diagonal. Integrating columns launched in the ground band determines its $Q\Pi$ block, and Hermiticity determines the reverse block. Equation~\eqref{eq:interband-cubic-coefficient} gives $\|K_3\|_F=2.76938\times10^{-10}$ at field rank 128.

Transform $R_{{\rm g},3}$ to the joint basis, including $e^{-ie_0T}$, and denote the result by $\mathcal B(R_{{\rm g},3})$. The two leading local defects of the composite operator are
\begin{equation}
\begin{aligned}
 E_{{\rm g},3}&=[\mathcal B(R_{{\rm g},3})\oplus0]
                          e^{-i(K_1+K_2)},\\
 E_{{\rm m},3}&=-i\widetilde F_{\rm d}K_3,\qquad
 E_3=E_{{\rm g},3}+E_{{\rm m},3}.
\end{aligned}
\label{eq:unified-local-defect}
\end{equation}
Here the two cubic indices refer respectively to the common ground-band response and to interband transitions. Equation~\eqref{eq:unified-local-defect} follows by expanding the difference of the exact and approximate ordered products; its coefficients are obtained independently of the long-sequence reference.

Propagate the leading residual together with the approximate state:
\begin{equation}
\begin{aligned}
 \psi_{n+1}&=\widetilde F\psi_n,\qquad \psi_0=\Psi_{\rm in},\\
 r_{n+1}&=\widetilde F r_n+E_3\psi_n,\qquad r_0=0,\\
 r_N&=\sum_{n=0}^{N-1}\widetilde F^{N-1-n}
                       E_3\widetilde F^n\Psi_{\rm in}.
\end{aligned}
\label{eq:accumulated-leading-residual}
\end{equation}
In a Floquet basis of $\widetilde F$, Eq.~\eqref{eq:accumulated-leading-residual} contains precisely the finite sum $D_N(\theta_\alpha-\theta_\beta)$ of Eq.~\eqref{eq:response-passage-sum}. This sum gives the resonant growth and off-resonant oscillations of the residual. The observable correction follows from the partial trace of $|r_N\rangle\langle\psi_N|+|\psi_N\rangle\langle r_N|$; the numerical reconstruction also retains $|r_N\rangle\langle r_N|$.

Over the 81 passages, the ground cubic defect contributes at most $1.23\times10^{-11}$ to $P_2$ and $7.31\times10^{-11}$ to complex $c_2$. The corresponding interband cubic contributions are about $9\times10^{-15}$ and $1.20\times10^{-13}$, respectively. The first value is on the integration error scale. In the employed Floquet basis, the diagonal part of the ground defect contributes a coherence correction of up to $5.66\times10^{-9}$. Adding the near-resonant off-diagonal terms reduces the maximum complete correction to $7.31\times10^{-11}$.

\begin{figure*}[t]
\centering
\includegraphics[width=\textwidth]{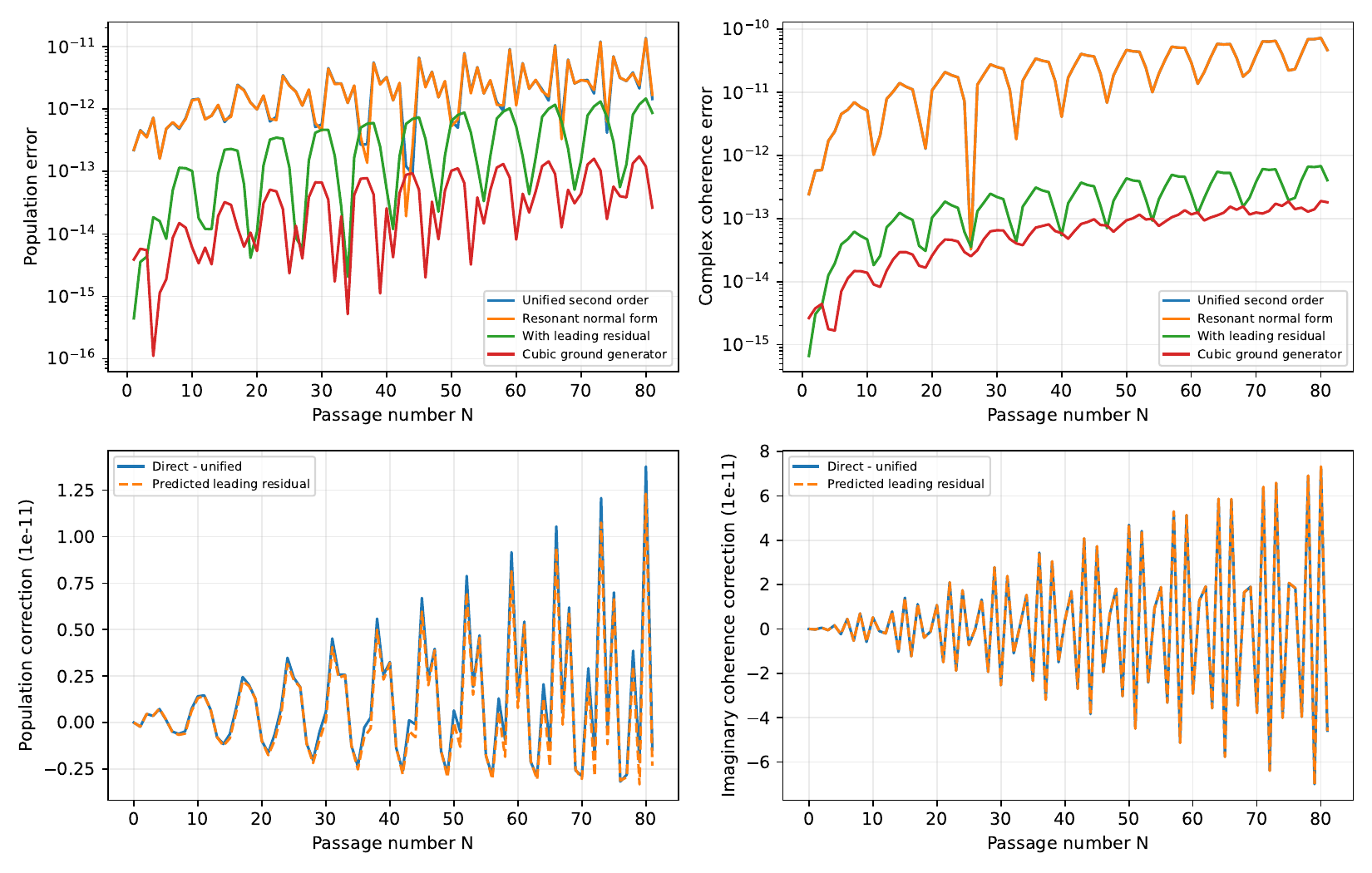}
\caption{Unified passage approximation and its coefficient-derived correction. Upper panels: errors in population and complex coherence, including the resonant normal form with nonresonant dressing. Lower panels: signed differences from the direct reference and the independently predicted leading residual. The cubic ground generator reduces the observable errors to about $2\times10^{-13}$ in this finite representation.}
\label{fig:unified-modulation}
\end{figure*}

The ground cubic coefficient can itself be resummed. Define $G_3$ by
\[
 (G_3)_{\alpha\beta}
 =\frac{(W_0^\dagger R_{{\rm g},3}W_0)_{\alpha\beta}}
 {f[\theta_\alpha,\theta_\beta]},
\]
and replace $G_{\rm ad}$ by $G_{\rm ad}+G_3$ in Eq.~\eqref{eq:unified-passage}. Table~\ref{tab:unified-modulation} compares the original unified approximation, the predicted leading residual, and this cubic ground-band generator. Figure~\ref{fig:unified-modulation} shows the signed correction and the remaining errors throughout the sequence.

\begin{table*}[t]
\centering
\caption{Maximum errors relative to full rank-128 propagation on each passage window. Each entry lists the errors in $(P_2,c_2)$. All three constructions use the same field representation. The residual column uses Eq.~\eqref{eq:accumulated-leading-residual}; the last column exponentiates the computed ground cubic generator.}
\label{tab:unified-modulation}
\begin{tabular}{rccc}
\toprule
$N_{\max}$ & Unified quadratic & With predicted leading residual & Cubic ground generator\\
\midrule
27 & $(3.48\times10^{-12},2.10\times10^{-11})$ & $(3.48\times10^{-13},1.86\times10^{-13})$ & $(5.12\times10^{-14},4.66\times10^{-14})$\\
54 & $(7.89\times10^{-12},4.70\times10^{-11})$ & $(8.80\times10^{-13},4.34\times10^{-13})$ & $(1.12\times10^{-13},1.16\times10^{-13})$\\
81 & $(1.38\times10^{-11},7.30\times10^{-11})$ & $(1.47\times10^{-12},6.80\times10^{-13})$ & $(1.75\times10^{-13},1.90\times10^{-13})$\\
\bottomrule
\end{tabular}
\end{table*}

The unified quadratic generator has a maximum coherence error $7.30\times10^{-11}$ over 81 passages. Its leading coefficient predicts this accumulated discrepancy, and the cubic resummation reduces it to $1.90\times10^{-13}$. The resonant normal form changes the quadratic result by less than $3.2\times10^{-13}$ in either observable. These errors compare propagation in the same finite representation. Full-space and basis-refinement comparisons are given in Appendix~\ref{app:modulation-validation}.

\subsection{Physical interpretation and controlled accuracy}
The modulated quantities are the complex coefficients of the joint quantum state. Their phases and occupations carry the translational response and the amplitudes of the discrete ring modes, while the chain exchange transfers the endpoint perturbation to the internal qubit. Equations~\eqref{eq:unified-passage} and \eqref{eq:discrete-resonant-modulation} retain this coupled evolution between encounters. The ordered return term in Eq.~\eqref{eq:mode-return-generator} accounts for the leading oscillatory correction to the internal-qubit response. Separately, Eq.~\eqref{eq:accumulated-leading-residual} resolves the next ground-band response coefficient: compensation between its diagonal and off-diagonal terms determines the error remaining after quadratic response resummation.

\begin{figure*}[t]
\centering
\includegraphics[width=.82\textwidth]{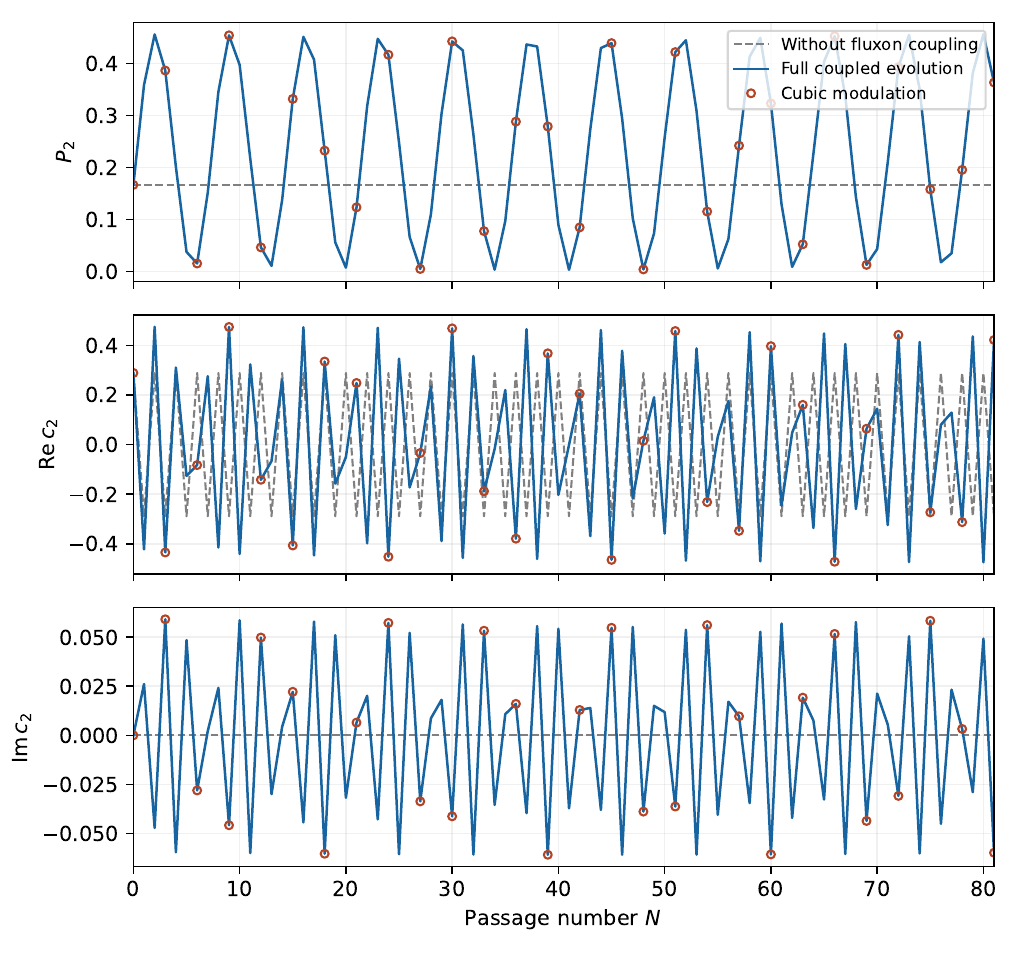}
\caption{Internal-qubit observables at the velocity-selected resonance. Solid curves show the full calculation with 24 translational states; open symbols, plotted every third passage, show the cubic ground-response modulation construction with dynamical excited modes. Dashed curves give exact chain evolution with $\kappa=0$, sampled at the same times $NT$. All curves use the chain preparation in Eq.~\eqref{eq:initial}. The coherence is expressed in the common-frequency rotating frame. Differences between the two coupled calculations are resolved in Fig.~\ref{fig:unified-modulation}.}
\label{fig:observable-overview}
\end{figure*}

Figure~\ref{fig:observable-overview} shows the population and the real and imaginary parts of the coherence themselves, together with the uncoupled-chain control and the reconstructed modulation solution. In the full calculation $P_2$ ranges from $0.00316$ to $0.45654$ over the sampled passages, compared with $P_2^{(0)}=1/6$. The differences of the mode-resolved and adiabatic calculations quantify the additional contribution from excited field branches on top of this driven response.

Direct propagation with all 86 retained oscillator states and 24 translational states gives a mode correction of approximately $6.1\times10^{-8}$ in population and complex coherence relative to the direct adiabatic ground-band control at the same parameters. The cubic ground-band approximation differs from this expanded reference by at most $1.5\times10^{-9}$. Increasing the translational basis from 16 to 24 changes these observables by less than $7.6\times10^{-12}$. An a posteriori estimate uses the independently integrated reference period map to evaluate the defect recurrence. After subtracting the calculated leading residual in the projected model, the resulting statewise bound at passage 81 is $2.78\times10^{-10}$. Combining it with the directly measured representation difference bounds the observable comparison errors by $2.05\times10^{-9}$ and $1.85\times10^{-9}$. Appendix~\ref{app:modulation-validation} separates the approximation, projection, and integration errors and derives the accumulated bound.

\section{Ordered radiative expansion}
\label{sec:harmonic-response}
The harmonic collective Hamiltonian in Eq.~\eqref{eq:hamiltonian} makes the emission and return processes explicit mode by mode. We now use its harmonic-vacuum initial condition and expand in the oscillator-source parameter $\lambda$, retaining the translational and chain evolution in $H_0$ exactly. This construction gives the operator memory kernel and the mixed-overlap diagnostic used in the subsequent mass and frequency comparisons. In the nonlinear modulation calculation above, the analogous return processes run through the excited eigenstates of the matched field Hamiltonian.
\subsection{Emission and return amplitudes}
Expand Eq.~\eqref{eq:system} in $\lambda$ while propagating $H_0$ exactly:
\begin{equation}
 \Psi=A|\boldsymbol0\rangle+\lambda\sum_\nu B_\nu|1_\nu\rangle
       +\lambda^2\sum_\nu C_\nu|\boldsymbol0\rangle
       +\lambda^2\Psi_{2\mathrm q}+\cdots .
 \label{eq:expansion}
\end{equation}
Here $A$, $B_\nu$, and $C_\nu$ are chain--packet vectors: the leading vacuum amplitude, the one-quantum emission amplitude, and its vacuum-return correction. The vector $\Psi_{2\mathrm q}$ collects two-quantum amplitudes. The coefficients needed for the density matrix through order $\lambda^2$ satisfy
\begin{align}
 i\dot A&=H_0A,\nonumber\\
 i\dot B_\nu&=(H_0+\omega_\nu)B_\nu+h_\nu Z_1A,\nonumber\\
 i\dot C_\nu&=H_0C_\nu+h_\nu^*Z_1B_\nu.
 \label{eq:hierarchy}
\end{align}
Their initial data follow from Eq.~\eqref{eq:initial}: $A(0)=\varphi c_{\rm in}$ and $B_\nu(0)=C_\nu(0)=0$. Orthogonality of Fock sectors in Eq.~\eqref{eq:expansion} gives
\begin{align}
 \rho&=\rho_X+\lambda^2\sum_\nu\Delta\rho_\nu+O(\lambda^4),\nonumber\\
 \rho_X&=\int AA^\dagger\dd\xi,\nonumber\\
 \Delta\rho_\nu&=\int\left(B_\nu B_\nu^\dagger
                   +C_\nu A^\dagger+AC_\nu^\dagger\right)\dd\xi.
 \label{eq:density-expansion}
\end{align}
The return amplitudes in Eq.~\eqref{eq:hierarchy} supply the interference terms in Eq.~\eqref{eq:density-expansion} and enforce trace preservation.

For a base density matrix with nonzero relevant entries, define
\begin{equation}
 \mathcal L_{\rho,j}(r)=-\Re\left[
 \frac{r_{j0}}{\rho_{j0}}-\frac{r_{jj}}{2\rho_{jj}}
                         -\frac{r_{00}}{2\rho_{00}}\right].
 \label{eq:functional}
\end{equation}
Expansion of Eq.~\eqref{eq:observable} then yields the complete second-order mixed response
\begin{equation}
 \Gamma_{{\rm mix},j}^{(2)}=
 \sum_\nu\left[\mathcal L_{\rho_X,j}(\Delta\rho_{X,\nu})
              -\mathcal L_{\rho_R^{(0)},j}(\Delta\rho_{R,\nu})\right].
 \label{eq:mixed-order}
\end{equation}
Here $\rho_R^{(0)}$ is the pure chain evolution on the prescribed trajectory, and the superscript denotes the coefficient of $\lambda^2$. The numerical evaluation of Eq.~\eqref{eq:mixed-order} includes every term in Eq.~\eqref{eq:density-expansion}.

\subsection{Operator memory kernel}
Let $U_0(t)$ propagate $H_0(t)$ on the chain and packet space, and set
\begin{align}
 S_\nu(t)&=U_0^\dagger(t)h_\nu(t,\xi)Z_1U_0(t),\nonumber\\
 Q_\nu(t)&=\int_0^t e^{i\omega_\nu s}S_\nu(s)\dd s,\nonumber\\
 R_\nu(t)&=\int_0^t\dd s\int_0^s\dd u\,
 e^{-i\omega_\nu(s-u)}S_\nu^\dagger(s)S_\nu(u).
 \label{eq:kernel}
\end{align}
For the initial chain--packet density operator $\varrho_{\mathrm{in}}=|\varphi c_{\rm in}\rangle\langle\varphi c_{\rm in}|$, the interaction-picture correction is
\begin{equation}
 \Delta\varrho_I=\sum_\nu
 (Q_\nu\varrho_{\mathrm{in}}Q_\nu^\dagger
       -R_\nu\varrho_{\mathrm{in}}-\varrho_{\mathrm{in}}R_\nu^\dagger).
 \label{eq:channel}
\end{equation}
Transformation by $U_0$ and tracing the packet in Eq.~\eqref{eq:channel} reproduces Eq.~\eqref{eq:density-expansion}. Splitting the time square into its two triangles gives
\[
 R_\nu+R_\nu^\dagger=Q_\nu^\dagger Q_\nu.
\]
Consequently the correction separates into the dissipator of $Q_\nu$ and the commutator with $(R_\nu-R_\nu^\dagger)/(2i)$. The ordered kernel in Eq.~\eqref{eq:kernel} retains the memory of both the packet and the chain over the entire observation time.

\section{Translation, curvature, and finite mass}
For the translational problem at $\lambda=0$, define $H_c(t,\xi)=H_q-(\kappa/\hstar)J(X_{\rm cl}+\xi)Z_1$. For a frozen displacement $\xi$, let $U(t)$ be the chain propagator at $\xi=0$, $\mathsf U_N=U(T)^N$, and $W(t,\xi)$ the propagator at that displacement. Since $H_c(t,\xi)=H_c(t+\xi/v,0)$ in this frozen problem,
\begin{equation}
\begin{aligned}
 W(t,\xi)&=U(t+\xi/v)U^\dagger(\xi/v),\\
 W(NT,\xi)&=U(\xi/v)\mathsf U_NU^\dagger(\xi/v).
\end{aligned}
\label{eq:conjugation}
\end{equation}
Differentiating Eq.~\eqref{eq:conjugation} with $H_b=H_c(0,0)$ gives
\begin{align}
 W_\xi(NT,0)&=-\frac{i}{v}[H_b,\mathsf U_N],\nonumber\\
 W_{\xi\xi}(NT,0)&=-\frac{i}{v^2}[\dot H_b,\mathsf U_N]
                    -\frac{1}{v^2}[H_b,[H_b,\mathsf U_N]].
 \label{eq:curvature-identity}
\end{align}
Equation~\eqref{eq:curvature-identity} constrains the sum of the direct $J''$ vertex and the iterated $J'$ vertices. Their compensation leaves the phase-of-entry dependence of the conjugated map. With $\rho_{\mathrm{in}}=c_{\rm in}c_{\rm in}^\dagger$, the frozen-packet density is the average of $W\rho_{\mathrm{in}}W^\dagger$ with weight $|\varphi|^2$.

For example, put $c=\mathsf U_Nc_{\rm in}$ and $d_r=\partial_\xi^rW(NT,0)c_{\rm in}$. For a narrow Gaussian packet,
\[
 1-\eta_j=\frac{\sigma_X^2}{2}
       \left|\frac{d_{1,j}}{c_j}-\frac{d_{1,0}}{c_0}\right|^2
       +O(\sigma_X^4).
\]
When the first conditional derivative difference vanishes exactly, the leading term is $\sigma_X^4|d_{2,j}/c_j-d_{2,0}/c_0|^2/4$. Quantitative results below use the unexpanded $J$.

Finite mass enters through an operator-valued connection. With $p=-i\hstar\partial_\xi$, substitute $A=W\chi$ and define
\begin{equation}
 \mathcal A=i\hstar W^\dagger\partial_\xi W,
 \qquad i\dot\chi=\frac{(p-\mathcal A)^2}{2M\hstar}\chi.
 \label{eq:connection}
\end{equation}
The first inverse-mass correction obtained from Eq.~\eqref{eq:connection} is
\begin{align}
 \Delta A=\frac{i\hstar}{2M}W
 \left[t\varphi''c_{\rm in}+2\varphi'\mathcal Q_1c_{\rm in}
                         +\varphi\mathcal Q_2c_{\rm in}\right],\nonumber\\
 \mathcal Q_1=\int_0^t W^\dagger W_\xi\dd s,\qquad
 \mathcal Q_2=\int_0^t W^\dagger W_{\xi\xi}\dd s.
 \label{eq:mass-expansion}
\end{align}
For real initial $\varphi$, the $t\varphi''$ term has zero first-order contribution to the reduced density matrix. The remaining terms depend on spatial derivatives of the chain propagator throughout the encounter sequence. Free spreading enters the variance as $\sigma_X^2+(\hstar t/(2M\sigma_X))^2$, while Eq.~\eqref{eq:mass-expansion} also retains the history-dependent backaction.

\section{Repeated passages and resonant phases}
Let $\mathsf U_T=U_0(T)$ act on the full chain--packet space. Periodicity implies $S_\nu(nT+s)=\mathsf U_T^{-n}S_\nu(s)\mathsf U_T^n$. Define $K_\nu=Q_\nu(T)$ and the one-period return $R_{\nu,1}$. Then
\begin{align}
 Q_{\nu,N}&=\sum_{n=0}^{N-1}q_{\nu,n},\qquad
 q_{\nu,n}=e^{i\omega_\nu nT}\mathsf U_T^{-n}K_\nu \mathsf U_T^n,\nonumber\\
 R_{\nu,N}&=\sum_{n=0}^{N-1}\mathsf U_T^{-n}R_{\nu,1}\mathsf U_T^n
                      +\sum_{n>m}q_{\nu,n}^\dagger q_{\nu,m}.
 \label{eq:passage}
\end{align}
Equation~\eqref{eq:passage} composes the emission and return amplitudes across passages.

In a spectral representation $\mathsf U_T|a\rangle=e^{-i\theta_a}|a\rangle$,
\begin{equation}
 (Q_{\nu,N})_{ab}=(K_\nu)_{ab}D_N(\Phi_{\nu,ab}),\quad
 \Phi_{\nu,ab}=\omega_\nu T+\theta_a-\theta_b,
 \label{eq:filter}
\end{equation}
where
\begin{equation}
 D_N(\Phi)=\sum_{n=0}^{N-1}e^{in\Phi}
 =e^{i(N-1)\Phi/2}\frac{\sin(N\Phi/2)}{\sin(\Phi/2)}.
 \label{eq:dirichlet}
\end{equation}
The squared phase sum obeys
\[
 |D_N(\Phi)|^2=N+2\sum_{\ell=1}^{N-1}(N-\ell)\cos(\ell\Phi).
\]
Thus $|D_N(\Phi)|^2/N$ is the Fej\'er kernel. The same triangular harmonic weights occur in the quantum-phase-estimation distribution for coherence-transfer coefficients that depend on the difference of register indices~\cite{gordeichuk2026qpe}. Here $\ell$ counts the separation between passages, and the phase $\Phi$ contains the field frequency and the chain--packet quasiphase difference in Eq.~\eqref{eq:filter}. Averaging over packet-dependent phases weights the interpassage correlations, as illustrated in Appendix~\ref{app:kinematic-memory}.

Continuous parts of the chain--packet Floquet spectrum are integrated with their spectral measures; the numerical packet basis gives a finite matrix representation. For returns, the additional finite sum is
\begin{equation}
 E_N(A,B)=\sum_{n>m}e^{-inA+imB}
       =\frac{D_N(-A)-D_N(B-A)}{1-e^{iB}},
 \label{eq:ordered-sum}
\end{equation}
with coincident-phase values defined by the finite sums in Eqs.~\eqref{eq:dirichlet} and \eqref{eq:ordered-sum}. Inserting Eq.~\eqref{eq:ordered-sum} into Eq.~\eqref{eq:passage} gives
\begin{align}
 (R_{\nu,N})_{ab}&=(R_{\nu,1})_{ab}D_N(\theta_a-\theta_b)\nonumber\\
 &\quad+\sum_c(K_\nu)_{ca}^*(K_\nu)_{cb}
              E_N(\Phi_{\nu,ca},\Phi_{\nu,cb}).
 \label{eq:return-filter}
\end{align}
Equations~\eqref{eq:filter} and \eqref{eq:return-filter}, followed by the functional in Eq.~\eqref{eq:functional}, determine the signed response.

Coherent accumulation of a contributing channel occurs for $|\Phi_{\nu,ab}-2\pi k|=O(N^{-1})$. Set $z=N(\Phi-2\pi k)$. The leading functions are
\begin{align}
 |D_N(z/N)|^2&\sim N^2 f(z),\quad
 f(z)=\left[\frac{\sin(z/2)}{z/2}\right]^2,\nonumber\\
 N^{-2}E_N(z/N,z/N)&\longrightarrow
       \frac{1-e^{-iz}-iz}{z^2}.
 \label{eq:resonance}
\end{align}
The real and imaginary parts of the second function in Eq.~\eqref{eq:resonance} are $f(z)/2$ and $-(z-\sin z)/z^2$. Their coefficients contain the source projections and the final conditional chain state. A sign boundary depends on the combination of direct and ordered terms.

The mean-chain Floquet phases for the parameters below are approximately $0,0,\pm0.400929$. They supply resonant offsets $0$, $\pm0.400929$, and $\pm0.801858$ in $\omega_\nu T$. Reconstructing the prescribed-trajectory response from one-period operators agrees with continuous integration of Eq.~\eqref{eq:hierarchy} to $1.2\times10^{-13}$ per mode over 36 passages.

\section{Source geometry and numerical results}
\subsection{Parameters and numerical methods}
The source in Eq.~\eqref{eq:projections} is a six-lobe periodic function. Its parameters were selected in a preliminary numerical search over source geometries using conditional overlaps at the endpoint and internal qubits. We keep this profile fixed in the mass, frequency, and mode-count comparisons below:
\begin{align}
 g(x)&=A_g\left[P_a(x)+rP_b(x)-(1+r)P_c(x)\right],\nonumber\\
 P_d(x)&=\frac{1}{\pi w}\sum_{m\in\mathbb Z}
 \biggl[\sech\frac{x+mL-d}{w}\nonumber\\ &\hspace{37mm}+\sech\frac{x+mL+d}{w}\biggr].
 \label{eq:source}
\end{align}
The integral of Eq.~\eqref{eq:source} over a period vanishes. We use
\begin{gather*}
 L=12,\quad v=0.54,\quad T=22.2222222,\quad \mathcal JT=0.4,\\
 \kappa=0.003,\quad \sigma_X=0.2,\\\hstar=5.96262690\times10^{-4},\\
 (a,b,c)=(0.92337631,3.37867757,3.93633430),\\
 w=0.90884831,\quad r=-1.64758581,\\
 A_g=0.42774526.
\end{gather*}
Equation~\eqref{eq:mass} gives $M_0=13.41599877$. These parameters specify the initial mass and frequency comparisons. Section~\ref{sec:resummed-modulation} gives the velocity-selected resonant parameters used for the long-sequence modulation test.

We discretize the ring with 192 Fourier collocation points and use its positive-frequency, positive-norm modes. The packet is represented by a Gaussian discrete-variable representation constructed from a truncated Hermite basis. Equation~\eqref{eq:hierarchy} is integrated with DOP853, relative tolerance $10^{-10}$, absolute tolerance $10^{-12}$, and maximum time step $0.12$. Direct Fock-space propagation of Eq.~\eqref{eq:system} uses tolerances $2\times10^{-12}$ and $2\times10^{-14}$, and maximum step $0.08$. The joint direct calculations retain up to two total oscillator quanta, with selected three-quantum checks. The harmonic comparisons in this section use the initial state in Eq.~\eqref{eq:initial}. The velocity-selected modulation test in Sec.~\ref{sec:resummed-modulation} uses its separately specified parameters and matched oscillator ground state.

Increasing the packet basis from 32 to 48 states changes the internal-qubit loss by about $10^{-13}$ in the tested 20-mode direct sequence. In a 14-mode check, increasing the total occupation cutoff from two to three changes it by at most $1.08\times10^{-9}$, or $0.027\%$. Adding modes 21--32 changes the second-order radiative increment by less than $9.5\times10^{-11}$ over 36 passages. The second-order radiative result differs from the 20-mode two-quantum direct result by at most $5.75\times10^{-9}$, or $0.112\%$ of the total loss in this sequence.

\subsection{Compensation and the residual loss}
\begin{figure*}
 \includegraphics[width=\textwidth]{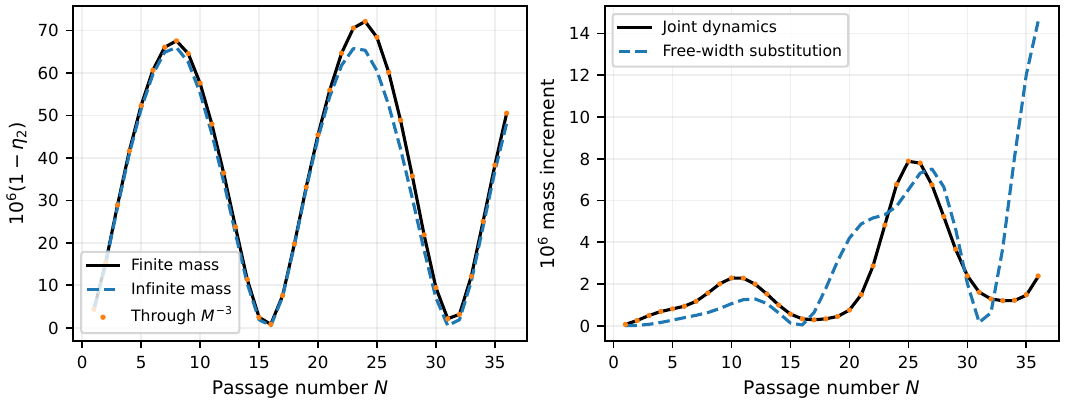}
 \caption{Translational contribution with the full position-dependent coupling. Left: infinite-mass and finite-mass results and the expansion through $M^{-3}$. Right: finite-mass increment compared with substituting the freely spread final width into the frozen-position formula.}
 \label{fig:mass}
\end{figure*}

Figure~\ref{fig:mass} compares the frozen-position average with finite-mass propagation. At passage 31 the translational loss rises from $5.09\times10^{-7}$ at infinite mass to $2.12\times10^{-6}$ at $M_0$. An inverse-mass expansion through $M^{-3}$ reproduces the internal-qubit translational loss to a maximum relative error of $0.24\%$ over the sequence. Replacing the initial width by its freely spread final value gives a different correction, as expected from the history terms in Eq.~\eqref{eq:mass-expansion}.

\begin{figure*}
 \includegraphics[width=\textwidth]{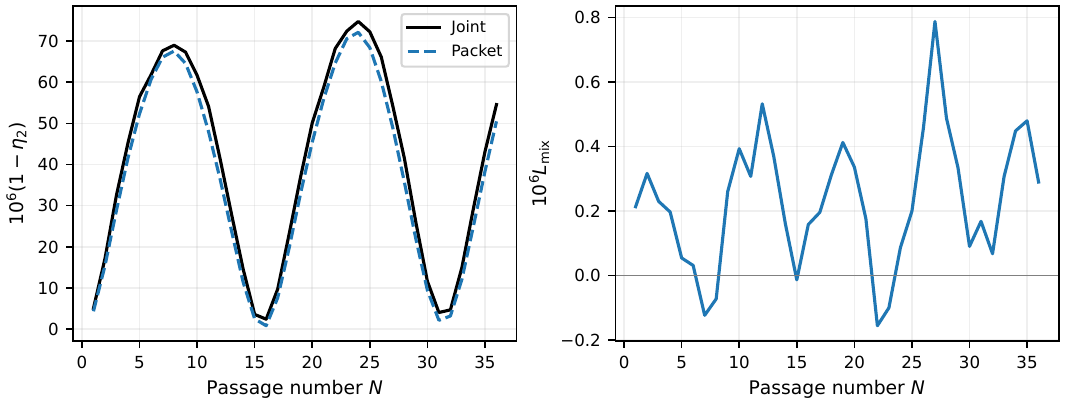}
 \caption{Joint loss and the signed mixed contribution for the internal qubit. The right panel resolves constructive and destructive mixed responses along the sequence of passages.}
 \label{fig:joint}
\end{figure*}

The joint calculation is shown in Fig.~\ref{fig:joint} and Table~\ref{tab:loss}. For this table $L=1-\eta_2$ and $L_{\rm mix}=L_{XR}-L_X-L_R$. The logarithmic definition in Eq.~\eqref{eq:mixed} removes the product correction associated with independent overlaps. At the displayed small losses the two signed diagnostics have the same signs.
\begin{table}
\caption{Direct conditional-overlap losses. Entries are multiplied by $10^6$; the joint calculation uses 20 modes, 48 packet states, and a two-quantum cutoff.}
\label{tab:loss}
\begin{ruledtabular}
\begin{tabular}{r r r r r}
$N$&$L_X$&$L_R$&$L_{\rm mix}$&$L_{XR}$\\
16&0.8028&1.4282&0.1577&2.3886\\
17&7.5796&1.9228&0.1955&9.6980\\
31&2.1191&1.7319&0.1676&4.0186\\
36&50.4895&3.7774&0.2895&54.5564
\end{tabular}
\end{ruledtabular}
\end{table}

The sign change can be resolved directly through Eq.~\eqref{eq:density-expansion}. At $N=22$, the change of the one-quantum part relative to the prescribed-trajectory calculation contributes $-1.09\times10^{-8}$ to $\Gamma_{\rm mix}^{(2)}$, while the vacuum-return part contributes $-1.44\times10^{-7}$. At $N=27$ the corresponding terms are $+7.63\times10^{-7}$ and $+2.57\times10^{-8}$. Vacuum-return interference supplies most of the suppression at passage 22, whereas the changed one-quantum contribution supplies most of the enhancement at passage 27.

\subsection{Sign boundaries under frequency and mass variation}
\begin{figure*}
 \includegraphics[width=\textwidth]{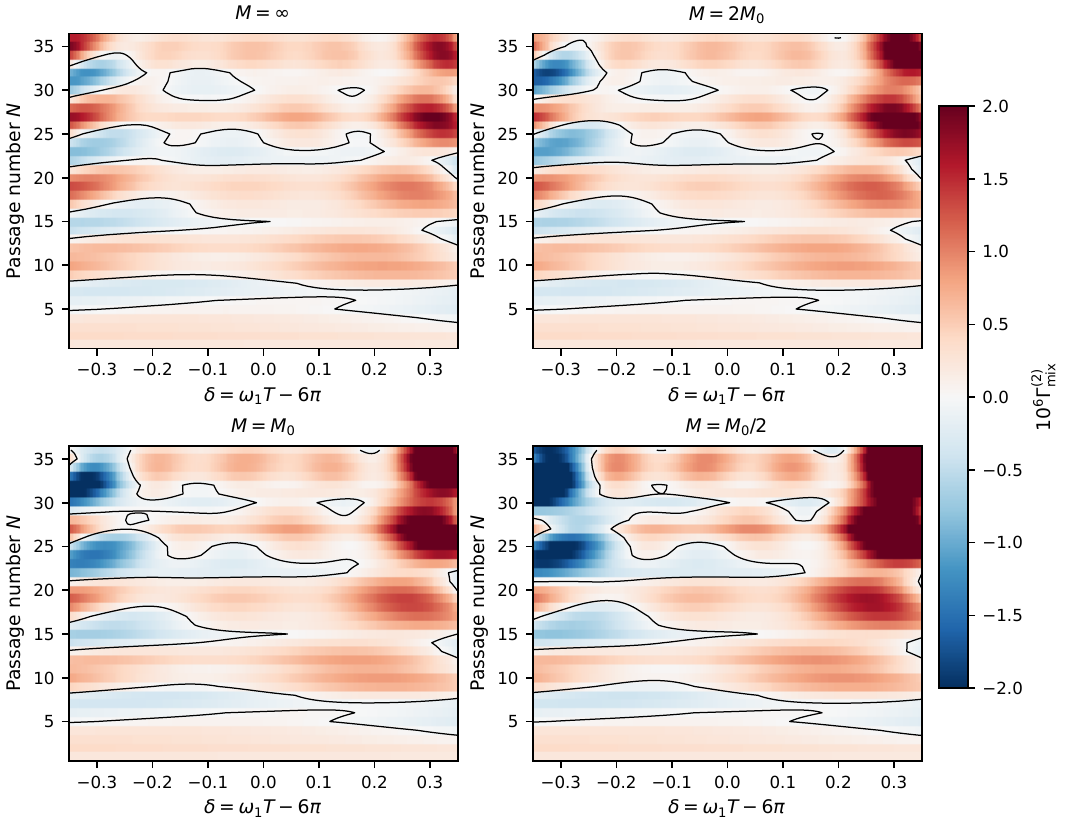}
 \caption{Mixed logarithmic response of the internal qubit. Blue denotes suppression and red enhancement relative to independent multiplication of the two overlaps. Black curves interpolate the zero boundaries between integer passages. The full second-order contribution includes all 20 modes and their vacuum returns.}
 \label{fig:map}
\end{figure*}

We vary $\delta=\omega_1T-6\pi$ while holding the source projections, the other frequencies, and $T$ fixed. The mass parameter is $\mu=M_0/M$. The resulting diagnostic separates frequency variation from source geometry. Figure~\ref{fig:map} shows $\Gamma_{\rm mix}^{(2)}$ for four masses. The frequency grid has spacing $0.005$ in $\delta$; an extended grid to $|\delta|=0.8$ closes intervals that leave the central plotting window.

At $N=18$, decreasing the mass from infinity to $M_0$ creates an interval $-0.23940<\delta<-0.17947$. At $M_0/2$ this interval becomes $-0.28580<\delta<-0.16725$. At $N=36$, the positive-detuning interval widens from $[0.19269,0.20480]$ at $2M_0$ to $[0.17423,0.20946]$ at $M_0$, and to $[0.15508,0.21002]$ at $M_0/2$. Its frequency widths are $0.000545$, $0.001585$, and $0.002472$.

At $N=27$ and $M=M_0/2$, returns produce a negative interval $[-0.31812,-0.20975]$. At $N=22$, the interval containing zero detuning joins a neighboring negative interval as mass decreases. Its width in frequency changes from $0.01620$ at infinite mass to $0.03358$ at $2M_0$, $0.03056$ at $M_0$, and $0.02829$ at $M_0/2$. The zero sets depend on both the resonant phase filters and the signed contributions of the remaining channels.

\subsection{Canonical center--mode correction}
The field reduction in Appendices~\ref{app:field-reduction} and \ref{app:canonical-vertices} yields the cubic center--mode energy~\eqref{eq:cm-vertices}. We evaluate its response at the physical mass $M=M_0$, holding the harmonic-mode parameters and source profile fixed. Normal ordering fixes the reference velocity by subtracting the vacuum coefficient of $p$. Direct propagation adds $\mathcal H_{3,\rm cm}/\hstar$ to Eq.~\eqref{eq:hamiltonian}; the translational control includes the same $c_3p^3$ term.

At $N=18$ and $\delta=-0.21$, the mixed logarithmic overlap changes from $-4.66018\times10^{-8}$ to $-2.12322\times10^{-7}$. The tangent equation~\eqref{eq:vertex-tangent} predicts $-2.15441\times10^{-7}$. Their absolute difference is $3.12\times10^{-9}$. The direct calculation uses 20 modes and total occupation cutoff two. Increasing the packet basis from 24 to 32 states changes the joint logarithmic overlap by $3.25\times10^{-12}$. The separately evaluated source-momentum vertex~\eqref{eq:source-momentum} contributes $4.70\times10^{-10}$ to the mixed response at first order. A joint refinement from 20 modes and 32 packet states to 24 modes and 24 packet states changes the tangent-corrected mixed response by $3.24\times10^{-12}$. In an eight-mode check, raising the total occupation cutoff from two to three changes the joint kinetic tangent by $2.30\times10^{-11}$ and the unperturbed logarithmic overlap by $8.28\times10^{-10}$. These comparisons resolve the cubic center--mode contribution at fixed harmonic dynamics. Table~\ref{tab:canonical-response} compares three detunings. The negative response at $\delta=-0.21$ persists, while $\delta=-0.26$ acquires a negative response and $\delta=-0.15$ remains positive.
\begin{table}
\caption{Mixed logarithmic overlap at $N=18$, $M=M_0$. Entries are multiplied by $10^7$. The last column is direct propagation with the cubic center--mode vertices at fixed reference velocity.}
\label{tab:canonical-response}
\begin{ruledtabular}
\begin{tabular}{r r r r}
$\delta$ & Harmonic & Linear correction & Direct cubic\\
$-0.26$ & $1.1061$ & $-1.2322$ & $-1.2168$\\
$-0.21$ & $-0.4660$ & $-2.1544$ & $-2.1232$\\
$-0.15$ & $1.1436$ & $0.6190$ & $0.6483$
\end{tabular}
\end{ruledtabular}
\end{table}

\section{Parameter dependence and spectral response}
The same emission and return mechanism produces the signed mixed response of Eq.~\eqref{eq:mixed}. In the harmonic collective model, changing the translational mass creates and merges intervals of reduced conditional-overlap loss. A local expansion at the folds gives square-root boundary laws. Appendix~\ref{app:mass-response} derives these laws and follows their changes when cubic and quartic field couplings are retained.

Matching excitation gaps and translational curvature permits a comparison across retained mode counts. At the diagnostic passage-27 reference point, the mixed response remains negative for 16, 20, and 24 modes. Spectral projections distinguish the field branches carrying excitation from those carrying conditional distinguishability: the dominant branches for these two weights differ. The ground-sector amplitudes, which retain the translational coordinate, carry most of the observed mode-count change in conditional distance. These diagnostics connect the coefficients of the joint amplitude evolution to the measured coherence response; their numerical values and convergence tests are collected in Appendix~\ref{app:mass-response}.

\section{Discussion and conclusions}
We have constructed modulation equations for the joint fluxon--chain amplitudes from the field-derived Hamiltonian and the one-passage variation hierarchy. The unitary passage approximation combines vacuum dressing, the translational packet, and excited-mode propagation. Its discrete resonant normal form evolves near-resonant amplitudes and restores the nonresonant components of the state. Population and complex coherence of the internal qubit follow by tracing the field degrees of freedom.

The generator at second order in interband transitions supplies the ordered-return correction accumulated during repeated encounters. The leading resonance produces population and coherence changes inside the chain, and the excited branches add the resolved mode contribution. At the next order in the ground-band response, diagonal and near-resonant off-diagonal terms partially compensate in the accumulated approximation error. This coefficient predicts the leading discrepancy and can itself be resummed. An exact residual recurrence, evaluated using an independently integrated reference map, gives an a posteriori state-dependent bound throughout the passage sequence.

At the velocity-selected resonance, the mode correction relative to adiabatic ground-band evolution is approximately $6.1\times10^{-8}$ in the tested observables. Direct full-space propagation over 81 passages places the error of the projected cubic approximation at $1.5\times10^{-9}$, with the 16-to-24-state translational refinement below $7.6\times10^{-12}$. The accumulated bound and the independently measured representation difference control the observable comparison at about $2\times10^{-9}$. The modulation solution reproduces the internal-qubit population and coherence through 81 passages with the stated errors.

\appendix
\section{Validation of the modulation approximation}
\label{app:modulation-validation}
\subsection{Accuracy over 81 passages}
The field construction uses 20 computed modes and total occupation at most two with energy bound 5. At this velocity, 18 one-quantum modes lie below that bound; the retained oscillator space has 86 states. A 12-state Hermite representation supplies the translational coordinate. The field projection includes the 12 ground-band states and additional orthogonal directions obtained from source-variation trajectories over the first two periods. We use total field ranks 64 and 128, with separately computed full-reference checks at intermediate ranks. The rank-128 chain--field space has dimension 512.

\begin{figure*}[t]
\centering
\includegraphics[width=\textwidth]{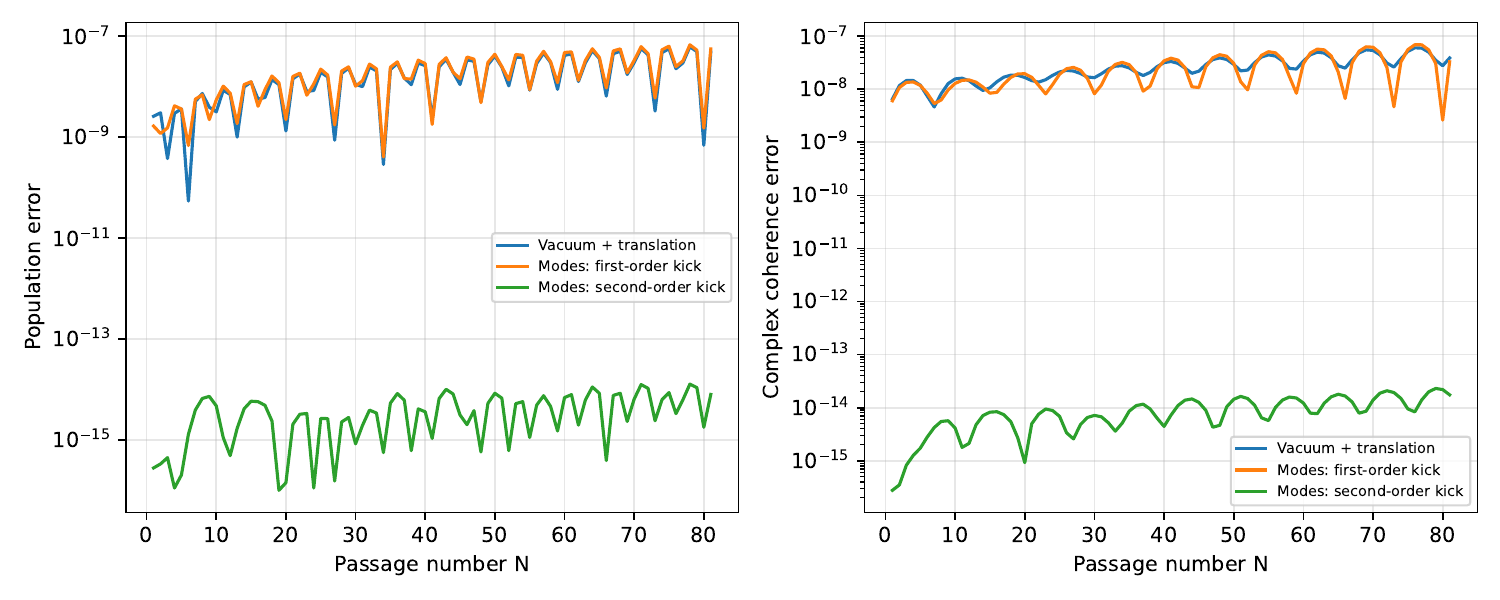}
\caption{Errors relative to full propagation in population (left) and complex coherence (right). The vacuum--translation generator leaves an accumulated mode correction. Inclusion of the first-order interband generator alone gives errors of comparable size. The ordered second-order term reproduces this correction; the green curves are on the numerical error scale. All comparisons use the fixed parameters of Sec.~\ref{sec:resummed-modulation}.}
\label{fig:resummed-81}
\end{figure*}

DOP853 integrates the period coefficient equations with relative tolerance $2\times10^{-13}$, absolute tolerance $2\times10^{-15}$, and maximum step $T/1024$. Direct integration of the full Hamiltonian supplies an independent reference period map, which is propagated for the same 81 passages. Table~\ref{tab:resummed-81} gives the maximum errors over every integer passage in each window. Figure~\ref{fig:resummed-81} resolves their passage dependence.

\begin{table*}[t]
\centering
\caption{Maximum absolute errors in population and complex coherence on $0\leq N\leq N_{\max}$. The full reference has field rank 128. The vacuum--translation generator uses 24 translational states; the dynamic-mode generator and reference use identical rank-128 field representations. The dynamic-mode comparison in the final two columns uses direct diagonal-block propagation and tests the interband truncation. These columns are on the numerical error scale established by step refinement; the unified approximation is tested in Table~\ref{tab:unified-modulation}.}
\label{tab:resummed-81}
\begin{tabular}{rrrrr}
\toprule
$N_{\max}$ & Vacuum--translation: $P_2$ & Vacuum--translation: $c_2$ & With $K_1+K_2$: $P_2$ & With $K_1+K_2$: $c_2$\\
\midrule
27 & $1.92\times10^{-8}$ & $2.26\times10^{-8}$ & $7.27\times10^{-15}$ & $9.43\times10^{-15}$\\
54 & $4.06\times10^{-8}$ & $4.04\times10^{-8}$ & $1.00\times10^{-14}$ & $1.63\times10^{-14}$\\
81 & $6.19\times10^{-8}$ & $6.01\times10^{-8}$ & $1.27\times10^{-14}$ & $2.30\times10^{-14}$\\
\bottomrule
\end{tabular}
\end{table*}

In the adiabatic model, the ordinary quadratic response has errors $1.31\times10^{-6}$ and $4.56\times10^{-7}$ by passage 81 in $P_2$ and $c_2$. Exponentiating the quadratic generator reduces them to $1.36\times10^{-11}$ and $7.31\times10^{-11}$. Relative to full field evolution, the residual is about $6\times10^{-8}$, as shown in Table~\ref{tab:resummed-81}. Retaining $K_1$ alone gives errors $6.77\times10^{-8}$ and $6.87\times10^{-8}$. Retaining $K_2$ as well reproduces the full result to the numerical integration scale. The rank-64 calculation has a maximum excited-band weight of approximately $1.49\times10^{-8}$; the ordered return amplitudes account for the observable correction despite this small weight.

Decreasing the maximum step from $T/1024$ to $T/4096$ at rank 64 changes $P_2$ and $c_2$ by at most $7.31\times10^{-14}$ and $5.42\times10^{-14}$. Expanding the field projection from rank 64 to 128 changes them by $1.54\times10^{-9}$ and $2.39\times10^{-9}$; the full-reference rank-96 to rank-128 changes are $2.38\times10^{-10}$ and $5.03\times10^{-10}$. These last values quantify the observed projection sensitivity. In the adiabatic calculation, increasing the translational basis from 16 to 24 changes the observables by $4.10\times10^{-12}$ and $7.44\times10^{-12}$. The tests separate the accuracy of the passage expansion, translational resolution, and field projection.

For an explicitly Hermitian generator we also replace the numerically integrated $F_{\rm d}$ by its unitary polar factor before taking the spectral logarithm. The resulting $F_{\rm eff}$ has unitarity residual $9.35\times10^{-14}$ in Frobenius norm, and reconstruction from $G_{\rm eff}$ has residual $4.83\times10^{-13}$. Its maximum observable differences from the direct rank-128 reference are $1.52\times10^{-14}$ and $4.14\times10^{-14}$. The agreement on the integration scale and the field-projection checks resolve the accumulated oscillatory correction throughout the 81-passage sequence.

\subsection{Normal-form refinement checks}
\label{app:normal-form-checks}
At the chosen mask width, the Frobenius norms are
\[
\begin{gathered}
 \|S_1\|_F=3.868\times10^{-4},\qquad
 \|S_2\|_F=1.153\times10^{-6},\\
 \|R_1\|_F=1.282\times10^{-7},\qquad
 \|R_2\|_F=3.414\times10^{-6}.
\end{gathered}
\]
There are 152 retained off-diagonal pairs with $|(k_1)_{\alpha\beta}|>10^{-12}$. The larger second-order norm reflects the accumulated return correction in this representation. Relative to $\widetilde F$, the maximum observable differences of $F_{\rm R}$ over 81 passages are $3.10\times10^{-13}$ in $P_2$ and $1.75\times10^{-13}$ in $c_2$. Halving or doubling the mask width keeps these differences below $3.14\times10^{-13}$ and $1.81\times10^{-13}$.

To check the expansion order, scale $k_1\mapsto xk_1$, $k_2\mapsto x^2k_2$, and use $xS_1,x^2S_2,xR_1,x^2R_2$ in Eq.~\eqref{eq:resonant-period-map}. For $x=1/4,1/2,1,2$, the Frobenius map error divided by $x^3$ lies between $2.76848\times10^{-10}$ and $2.76850\times10^{-10}$. This verifies the cubic remainder of the discrete normal-form transformation at fixed physical $\kappa$.

\subsection{Direct propagation with an enlarged translational basis}
\label{sec:full-translation-check}
We propagate the matched fourth-degree Hamiltonian directly in the tensor product of all 86 retained oscillator states, the translational Hermite basis, and the four chain states. The comparisons use $n_X=12,16,24$ throughout $0\leq t\leq81T$, with the same mode gaps, matched mass, source harmonics, and initial preparation. The $n_X=24$ calculation has 8256 complex amplitudes. This continuous integration supplies a reference independent of the rank-128 period-map projection.

The source action is evaluated by combining equal momentum blocks and applying their exact sparse Fock matrices. At three times within a period, the resulting right-hand side agrees with the original implementation to relative discrepancy $1.1\times10^{-16}$ on a normalized random state. DOP853 uses relative tolerance $2\times10^{-11}$, absolute tolerance $2\times10^{-13}$, and maximum step $0.5$. The refined 24-state run uses $2\times10^{-12}$, $2\times10^{-14}$, and $0.25$, respectively. Refinement changes $P_2$ by at most $5.33\times10^{-14}$ and $c_2$ by $6.59\times10^{-14}$ over the complete sequence. The refined norm drift is $1.13\times10^{-14}$.

Increasing $n_X$ from 16 to 24 changes $P_2$ by at most $4.32\times10^{-12}$ and $c_2$ by $7.53\times10^{-12}$. The maximum probability in the highest four retained Hermite states decreases from $6.25\times10^{-8}$ to $5.33\times10^{-12}$. The state vectors expressed in the common Hermite basis differ in norm by at most $2.02\times10^{-5}$; the small changes of the chain observables quantify their sensitivity to this field-state difference. Table~\ref{tab:full-basis-check} separates the translational change from the rank-128 joint-field projection at fixed $n_X=12$.

\begin{table*}[t]
\centering
\caption{Maximum observable differences over all 81 passages. Full propagation retains all 86 oscillator states. The rank comparison keeps $n_X=12$ fixed; the 24-state reference uses refined integration tolerances.}
\label{tab:full-basis-check}
\begin{tabular}{lrr}
\toprule
Comparison & $P_2$ & Complex $c_2$\\
\midrule
Full propagation: $n_X=12\to16$ & $7.00\times10^{-10}$ & $1.22\times10^{-9}$\\
Full propagation: $n_X=16\to24$ & $4.32\times10^{-12}$ & $7.53\times10^{-12}$\\
Rank 128 to full propagation at $n_X=12$ & $1.27\times10^{-9}$ & $8.60\times10^{-10}$\\
Rank 128 at $n_X=12$ to full propagation at $n_X=24$ & $1.49\times10^{-9}$ & $1.28\times10^{-9}$\\
Full $n_X=24$: integration refinement & $5.33\times10^{-14}$ & $6.59\times10^{-14}$\\
\bottomrule
\end{tabular}
\end{table*}

The change from 12 to 16 translational states reaches $1.22\times10^{-9}$ in coherence, while the next refinement reaches $7.53\times10^{-12}$. At fixed $n_X=12$, restoring the complete oscillator--translation space changes $P_2$ and $c_2$ by $1.27\times10^{-9}$ and $8.60\times10^{-10}$. Their combined effect relative to the full 24-state reference is $1.49\times10^{-9}$ and $1.28\times10^{-9}$. These comparisons resolve both contributions to the finite-representation error.

Relative to the direct 24-state adiabatic control, the full 24-state evolution gives a mode correction of $6.17\times10^{-8}$ in population and $6.09\times10^{-8}$ in complex coherence. The maximum errors of the cubic ground-generator approximation relative to this full reference are $1.49\times10^{-9}$ and $1.28\times10^{-9}$.

\subsection{Accumulated error after subtraction of the leading residual}
\label{sec:accumulated-error-bound}
Let $F$ be the full unitary period map in the chosen finite field representation, $A=\widetilde F$ the unified unitary approximation, and $E=E_3$ the computed leading local defect. Define
\[
 \Delta=F-A,\qquad R_{\rm loc}=\Delta-E.
\]
For a normalized initial state, propagate $\psi_n=A^n\Psi_{\rm in}$ and $r_n$ by Eq.~\eqref{eq:accumulated-leading-residual}. The error remaining after reconstruction is $q_n=F^n\Psi_{\rm in}-\psi_n-r_n$. Direct substitution gives the exact identity
\begin{equation}
 q_{n+1}=Fq_n+d_n,\qquad
 d_n=R_{\rm loc}\psi_n+\Delta r_n,\qquad q_0=0.
 \label{eq:remaining-error-recurrence}
\end{equation}
The first term in $d_n$ is the next local defect acting on the incident approximate state; the second accounts for repeated action of the map difference on the accumulated leading correction. Unitarity in Eq.~\eqref{eq:remaining-error-recurrence} gives the statewise bound
\begin{equation}
 \|q_N\|\leq B_N:=\sum_{n=0}^{N-1}\|d_n\|
 \leq\sum_{n=0}^{N-1}
 \bigl(\|R_{\rm loc}\psi_n\|+\|\Delta r_n\|\bigr).
 \label{eq:trajectory-error-bound}
\end{equation}
It retains the actual occupation of the propagated subspace. The corresponding uniform bound follows from $\|r_n\|\leq n\|E\|$:
\begin{equation}
 \|q_N\|\leq N\|R_{\rm loc}\|
      +\frac{N(N-1)}{2}\|\Delta\|\|E\|.
 \label{eq:uniform-error-bound}
\end{equation}
A sufficient condition for uniform accuracy is the smallness of both terms in Eq.~\eqref{eq:uniform-error-bound}. Equation~\eqref{eq:trajectory-error-bound} gives the sharper estimate for the specified preparation.

For either the population operator or the operator defining complex coherence, the operator norm is one. With $u_N=\psi_N+r_N$, the observable error satisfies
\begin{equation}
 \bigl|\langle O\rangle_{F^N\Psi_{\rm in}}
             -\langle O\rangle_{u_N}\bigr|
 \leq (1+\|u_N\|)B_N\leq 2B_N+B_N^2.
 \label{eq:observable-error-bound}
\end{equation}
The same bound applies to the complex off-diagonal element through $O=I_{\rm field}\otimes|2\rangle\langle0|$.

We evaluate these bounds using the polar unitary factors of independently integrated reference and approximate maps, with the same normalized projected initial state. The reference period is integrated with DOP853, relative tolerance $2\times10^{-13}$, absolute tolerance $2\times10^{-15}$, and maximum step $T/4096$. In spectral norm,
\[
\begin{gathered}
 \|\Delta\|=1.59240\times10^{-7},\qquad
 \|E\|=1.59226\times10^{-7},\\
 \|R_{\rm loc}\|=5.69083\times10^{-9}.
\end{gathered}
\]
At $N=81$, Eq.~\eqref{eq:uniform-error-bound} gives $4.61\times10^{-7}$, whereas the trajectory bound of Eq.~\eqref{eq:trajectory-error-bound} is $2.78\times10^{-10}$. The sum of repeated-insertion norms $\sum_n\|\Delta r_n\|$ is $7.45\times10^{-16}$; the next local defect determines the bound in this calculation. Table~\ref{tab:residual-bound} compares the bound with the directly measured state error on every passage window.

\begin{table}[t]
\centering
\caption{Accumulated error after subtraction of the leading residual in the rank-128 model. The measured error is the maximum of $\|q_n\|$ on $0\leq n\leq N$. The monotone bound $B_N$ controls this whole window.}
\label{tab:residual-bound}
\begin{tabular}{rrrr}
\toprule
$N$ & Measured error & $B_N$ & Observable bound\\
\midrule
27 & $7.28\times10^{-12}$ & $1.44\times10^{-11}$ & $2.89\times10^{-11}$\\
54 & $1.65\times10^{-11}$ & $7.48\times10^{-11}$ & $1.50\times10^{-10}$\\
81 & $3.44\times10^{-11}$ & $2.78\times10^{-10}$ & $5.57\times10^{-10}$\\
\bottomrule
\end{tabular}
\end{table}

Changing the reference maximum step from $T/2048$ to $T/4096$ changes the propagated reference state by at most $1.74\times10^{-13}$ and changes $B_{81}$ by $2.87\times10^{-5}$ relative. The observable errors after reconstruction are at most $1.08\times10^{-12}$ in $P_2$ and $5.16\times10^{-13}$ in $c_2$ for these unitary maps and normalized state.

\begin{figure*}[t]
\centering
\includegraphics[width=\textwidth]{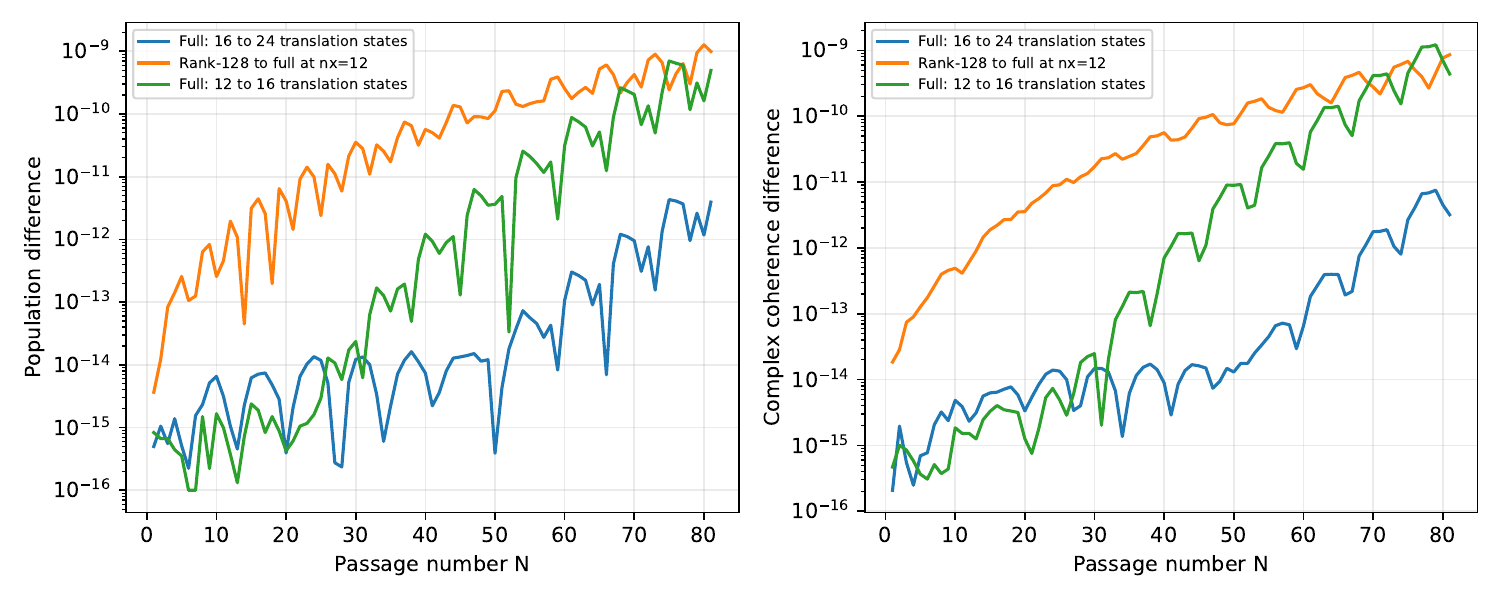}\\
\includegraphics[width=\textwidth]{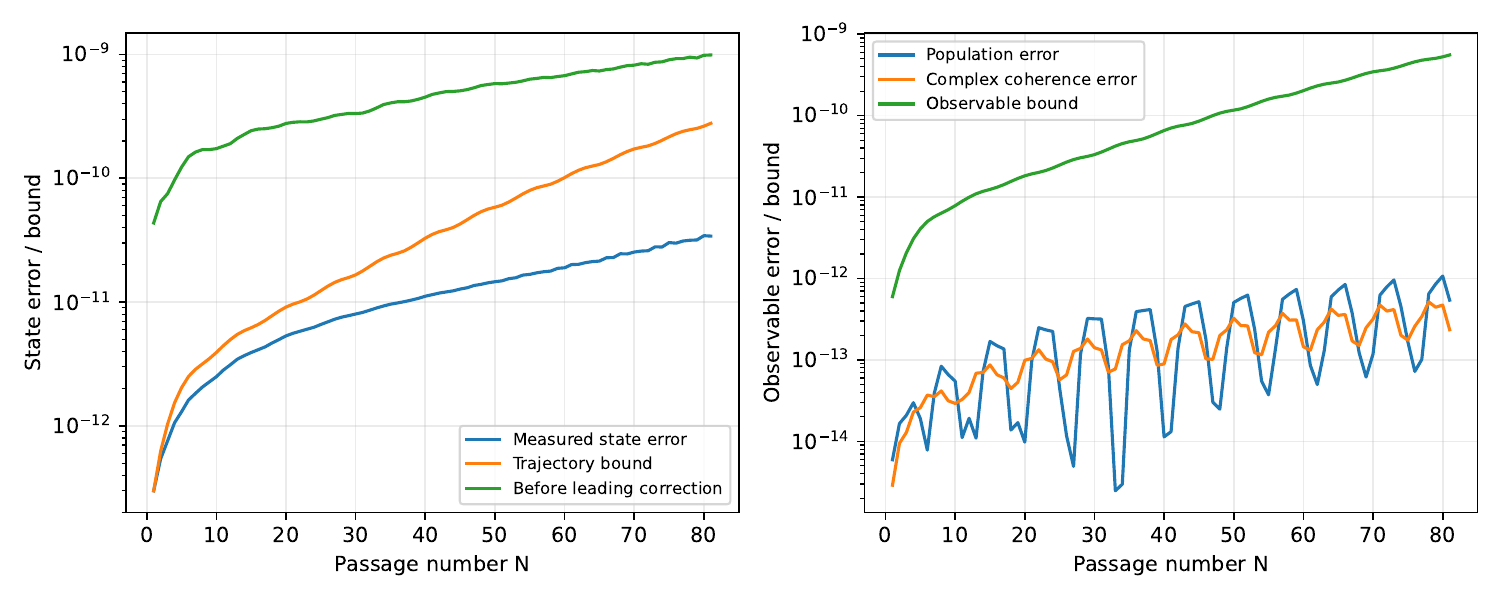}
\caption{Upper panels: direct translational-basis refinement and joint-field projection error, separately resolved in population and complex coherence. Lower panels: measured error after subtraction of the leading residual and the statewise bounds from Eqs.~\eqref{eq:trajectory-error-bound} and \eqref{eq:observable-error-bound}. The bounds are evaluated from the numerically constructed unitary maps at field rank 128; the full 24-state comparison is included through the separate observable representation error.}
\label{fig:full-basis-residual}
\end{figure*}

The representation comparison can be added at the observable level. Let $\eta_{O,N}$ be the maximum difference between the normalized rank-128 reference and the full refined 24-state reference on $0\leq n\leq N$. The triangle inequality and Eq.~\eqref{eq:observable-error-bound} give
\[
 \max_{n\leq N}|\langle O\rangle_{{\rm full},n}
                -\langle O\rangle_{u_n}|
 \leq\eta_{O,N}+2B_N+B_N^2.
\]
For $N=81$, the evaluated right-hand side is $2.05\times10^{-9}$ for $P_2$ and $1.85\times10^{-9}$ for $c_2$. These are a posteriori bounds for comparison with the explicitly tested full finite model. Figure~\ref{fig:full-basis-residual} displays the basis differences and the statewise remainder bound along the sequence.

\section{Mass-dependent response and spectral diagnostics}
\label{app:mass-response}
The calculations below use three successive Hamiltonians. The harmonic model gives the second-order radiative response and its local zero boundaries. The center--mode family adds the momentum-dependent cubic vertices and follows the boundaries as their strength increases. The fourth-degree model includes the pure-mode vertices and transformed source as well; its final comparison matches excitation gaps and translational curvature across mode counts. Each stage uses the joint evolution and its own two controls in Eq.~\eqref{eq:mixed}.

Write $\mathcal F(\delta,\mu;N)=\Gamma_{\rm mix}^{(2)}$. A critical point satisfies
\begin{equation}
 \mathcal F(\delta_c,\mu_c;N)=0,\qquad
 \mathcal F_\delta(\delta_c,\mu_c;N)=0.
 \label{eq:critical}
\end{equation}
We compute the local derivatives
\[
 A=\mathcal F_\mu,\quad B=\mathcal F_{\delta\delta},\quad
 C=\mathcal F_{\mu\delta},\quad D=\mathcal F_{\delta\delta\delta}
\]
in Eq.~\eqref{eq:critical}. With $d=\mu-\mu_c$ and $y=\delta-\delta_c$,
\begin{equation}
 \mathcal F=Ad+\tfrac12By^2+Cd y+\tfrac16Dy^3
                 +O(d^2,d y^2,y^4).
 \label{eq:normal-form}
\end{equation}
For $AB\ne0$, substitution into Eq.~\eqref{eq:normal-form} yields
\begin{equation}
\begin{aligned}
 \delta_\pm&=\delta_c+bd
       \pm\sqrt{-2Ad/B}+O(|d|^{3/2}),\\
 b&=-\frac CB+\frac{AD}{3B^2}.
\end{aligned}
\label{eq:boundaries}
\end{equation}
All four computed points have $A<0$. With $B>0$, a negative interval is born for $d>0$. With $B<0$, the roots at $d<0$ enclose a positive gap that closes as two negative intervals merge. These are folds of the zero level of the conditional response.

Tables~\ref{tab:critical} and \ref{tab:fold-coeff} give the critical parameters and coefficients in Eq.~\eqref{eq:boundaries}. The numerical surfaces are computed directly on a three-point mass grid and nine-point detuning grid. We evaluate the derivatives using a local quadratic fit in mass and a sixth-degree fit in detuning. Halving the mass spacing from $0.01$ to $0.005$ and reducing the detuning spacing from $0.0025$ to $0.0015$ change $\mu_c$ by less than $4\times10^{-10}$ in this calculation.
\begin{table}
\caption{Critical points of the full second-order mixed response. Births and the merger occur as $M$ decreases.}
\label{tab:critical}
\begin{ruledtabular}
\begin{tabular}{r l r r r}
$N$&Type&$\mu_c$&$\delta_c$&$M_c/M_0$\\
18&Birth&0.655623&$-0.201975$&1.525266\\
22&Merger&0.143801&$-0.210282$&6.954059\\
27&Birth&1.022757&$-0.247415$&0.977750\\
36&Birth&0.434304&$0.199826$&2.302532
\end{tabular}
\end{ruledtabular}
\end{table}
\begin{table}
\caption{Coefficients for Eq.~\eqref{eq:boundaries}. Here $s=\sqrt{|2A/B|}$ and the square-root term is $s\sqrt{|d|}$.}
\label{tab:fold-coeff}
\begin{ruledtabular}
\begin{tabular}{r r r r r}
$N$&$10^7A$&$10^4B$&$s$&$b$\\
18&$-1.329057$&1.046369&0.050402&$-0.022899$\\
22&$-2.417149$&$-0.763155$&0.079590&0.025675\\
27&$-5.409696$&2.716976&0.063104&$-0.023937$\\
36&$-1.276637$&4.587744&0.023591&$-0.016828$
\end{tabular}
\end{ruledtabular}
\end{table}

The width in frequency is $(\delta_+-\delta_-)/T$. For a birth, $M<M_c$ and $d=\mu_c(M_c-M)/M_c+O((M-M_c)^2)$, so
\begin{equation}
 \Delta\omega\sim\frac{2s\sqrt{\mu_c}}{T}
                  \sqrt{\frac{M_c-M}{M_c}}.
 \label{eq:mass-width}
\end{equation}
For the merger, replacing $M_c-M$ by $M-M_c$ in Eq.~\eqref{eq:mass-width} gives the width of the remaining positive gap. Independent integrations at $|d|=0.0025,0.01,0.04$ confirm the boundaries in Fig.~\ref{fig:folds}. The largest relative width errors over these points are $0.20\%$, $0.017\%$, $0.48\%$, and $0.11\%$ for $N=18,22,27,36$, respectively.
\begin{figure*}
 \includegraphics[width=\textwidth]{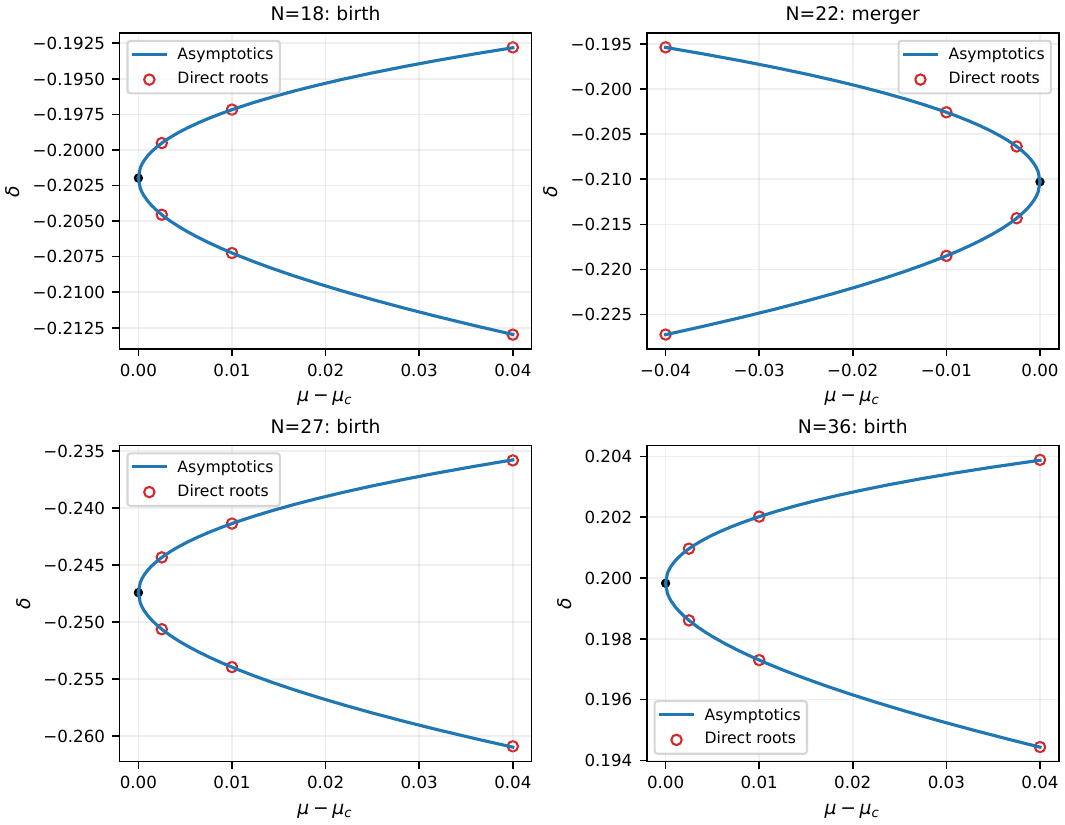}
 \caption{Local boundary expansions and independent integrations of the full second-order hierarchy. Red circles are roots inferred from direct evaluations on both sides of each boundary; the lines use Eq.~\eqref{eq:boundaries}. The merger closes a positive gap, while the other three panels show the birth of negative intervals.}
 \label{fig:folds}
\end{figure*}

Higher radiative orders shift the critical parameters. If $\Gamma_{\rm mix}=\lambda^2\mathcal F_2+\lambda^4\mathcal F_4+\cdots$, continuation of Eq.~\eqref{eq:critical} gives
\begin{equation}
 \mu_c^{(4)}=-\frac{\mathcal F_4}{A},\qquad
 \delta_c^{(4)}=-\frac{\partial_\delta\mathcal F_4+C\mu_c^{(4)}}{B}.
 \label{eq:critical-shift}
\end{equation}
A direct 20-mode calculation for $N=18$, with the joint occupation cutoff two and a three-quantum prescribed-trajectory calculation, gives a local control estimate $\mu_c\simeq0.67034$, $\delta_c\simeq-0.20134$, or $M_c\simeq1.492M_0$. This estimate uses direct frequency curvature and the second-order mass derivatives for the final correction in Eq.~\eqref{eq:critical-shift}. The mass shift is approximately $-2.2\%$; the local curvature remains positive. Tables~\ref{tab:critical} and \ref{tab:fold-coeff} specify the asymptotic critical points whose boundary expansions were tested above.

\subsection{Continuation of the center--mode cubic response}
To continue the corrected response in inverse mass, define the diagnostic family in energy units
\begin{equation}
 \mathcal H_{\mu,\epsilon}=\mathcal H_{\mathrm{harm},1}
 +\frac{\mu-1}{2M_0}p^2+\epsilon\mathcal H_{3,\rm cm},
 \qquad \mu=M_0/M.
 \label{eq:cubic-family}
\end{equation}
Here $\mathcal H_{\mathrm{harm},1}$ is the harmonic Hamiltonian in energy units at $M=M_0$. Equation~\eqref{eq:cubic-family} holds the source, ring modes, and cubic coefficients at their values for $v=0.54$. The parameter $\mu$ changes the quadratic translational curvature; $\epsilon$ controls the cubic vertices. Both the joint and translational-control propagators use this same prescription. At $\mu=1$, $\epsilon=1$, it reproduces the canonical-response check above.

Let $F_N(\delta,\mu,\epsilon)$ denote the directly propagated mixed logarithmic overlap. A fold satisfies $F_N=\partial_\delta F_N=0$. At a nondegenerate fold, differentiation with respect to $\epsilon$ gives
\[
 \frac{\dd\mu_c}{\dd\epsilon}=-\frac{F_\epsilon}{F_\mu},\qquad
 \frac{\dd\delta_c}{\dd\epsilon}
 =-\frac{F_{\delta\epsilon}+F_{\delta\mu}\,\dd\mu_c/\dd\epsilon}
          {F_{\delta\delta}}.
\]
All derivatives are evaluated on the fold. These equations establish local continuation while $F_\mu F_{\delta\delta}\ne0$. The relative sensitivity of the critical mass is
\begin{equation}
 \frac{1}{M_c}\frac{\dd M_c}{\dd\epsilon}
 =-\frac{1}{\mu_c}\frac{\dd\mu_c}{\dd\epsilon}
 =\frac{F_\epsilon}{\mu_c F_\mu}.
 \label{eq:mass-sensitivity}
\end{equation}
Equation~\eqref{eq:mass-sensitivity} separates the response sensitivity $F_\epsilon/F_\mu$ from the amplification by $1/\mu_c$. For a change of numerical truncation producing errors $\Delta F$ and $\Delta F_\delta$ at a fold, the first-order displacement is
\[
 \Delta\mu_c=-\frac{\Delta F}{A},\qquad
 \Delta\delta_c=-\frac{\Delta F_\delta+C\Delta\mu_c}{B},
\]
Here $A=F_\mu$, $B=F_{\delta\delta}$, and $C=F_{\delta\mu}$. These expressions relate response errors to shifts of the critical parameters. At $\epsilon=1$ in the center--mode family, the two zero conditions are solved using direct propagation of Eq.~\eqref{eq:cubic-family}. The resulting local boundaries have the form of Eq.~\eqref{eq:boundaries}, with derivatives of this corrected response.

For $N=18$ and $\mu=1$, direct cubic propagation brackets the tracked boundaries as
\[
\begin{aligned}
 -0.278658&<\delta_-<-0.277000,\\
 -0.161000&<\delta_+<-0.159995.
\end{aligned}
\]
At the left bracket endpoints, the mixed responses are $9.87\times10^{-9}$ and $-3.86\times10^{-9}$; at the right endpoints they are $-2.60\times10^{-9}$ and $3.43\times10^{-9}$. The detuning width is between $0.11600$ and $0.11866$, compared with $0.05993$ for the harmonic second-order interval quoted above. Linear interpolation gives approximate boundary locations $-0.27747$ and $-0.16057$.

The mass continuation for $N=18$ and $N=22$ reaches $\mu=0.005$ with negative responses near the tracked minimum and maximum. Direct 24-state packet calculations at the diagnostic boundary $\mu=0$ give $F_{18}(-0.2022483,0,1)=-8.14\times10^{-8}$ and $F_{22}(-0.175331502,0,1)=-6.59\times10^{-8}$. These values characterize the continued local branches at the boundary of the positive inverse-mass domain.

For $N=27$ and $N=36$, the corrected response has the nondegenerate folds in Table~\ref{tab:cubic-folds}. Both satisfy $A=F_\mu<0$ and $B=F_{\delta\delta}>0$. Consequently, increasing $\mu$ through $\mu_c$ creates a negative interval with local boundaries
\[
 \delta_\pm=\delta_c\pm s\sqrt{\mu-\mu_c}
                 +O(\mu-\mu_c),\qquad s=\sqrt{-2A/B}.
\]
The corresponding transition is traversed by decreasing $M$. Direct propagation uses 20 ring modes, a joint occupation cutoff of two, and 24 translational states. The relative and absolute integration tolerances are $2\times10^{-12}$ and $2\times10^{-14}$, with maximum time step $0.08$ in the dimensionless time variable. Phase derivatives use a centered difference with step $0.001$, and the inverse-mass derivative uses a forward step $0.01$. At the tabulated points, $|F|\le2.0\times10^{-10}$ and $|F_\delta|\le5.4\times10^{-9}$. The mass values are rounded to reflect these residuals.

\begin{table}[tb]
\caption{Folds of the directly propagated center--mode cubic response at occupation cutoff two, compared with the second-order harmonic critical masses. The inverse-mass family is defined by Eq.~\eqref{eq:cubic-family}; all cubic coefficients are fixed at their reference values.}
\label{tab:cubic-folds}
\begin{ruledtabular}
\begin{tabular}{rrrrr}
$N$ & $M_c/M_0$ (harm.) & $M_c/M_0$ (cubic) & $\delta_c$ & $s$\\
27 & 0.978 & 15.8 & $-0.24095$ & 0.0664\\
36 & 2.303 & 3.0 & $0.19545$ & 0.0206\\
\end{tabular}
\end{ruledtabular}
\end{table}

Independent direct propagations test the signs around each fold. For $N=27$, take $e=0.01$; for $N=36$, take $e=0.02$. In the ordered set of points
\[
 (\mu_c-e,\delta_c),\quad(\mu_c+e,\delta_c),\quad
 (\mu_c+e,\delta_c\pm2s\sqrt e),
\]
the response values, in units of $10^{-9}$, are respectively
\[
\begin{aligned}
 N=27:&\quad (5.46,-5.45,16.18,16.25),\\
 N=36:&\quad (2.83,-2.43,6.93,9.45).
\end{aligned}
\]
Both corrected folds retain the birth of a negative interval: the center changes from positive to negative, and the two outer points remain positive. The comparison in Table~\ref{tab:cubic-folds} includes both the cubic center--momentum contribution and the change from the second-order radiative hierarchy to direct occupation-cutoff propagation.

\subsection{Continuation in cubic strength and numerical sensitivity}
For passage 27, direct propagation at $\epsilon=0$ locates the harmonic fold at $\mu_c=1.03173988$, $\delta_c=-0.24727472$, or $M_c=0.96924M_0$. The same propagator at $\epsilon=1$ gives the cubic critical point in Table~\ref{tab:cubic-folds}. This comparison fixes the radiative occupation cutoff and the propagation method at both endpoints.

Table~\ref{tab:epsilon-fold} gives the direct continuation in cubic strength. At all five sampled points, $A<0$ and $B>0$. The magnitude of $A$ varies by about $2.1\%$, while $B$ decreases smoothly from $2.75\times10^{-4}$ to $2.47\times10^{-4}$. The determinant $AB$ remains between $-1.48\times10^{-10}$ and $-1.34\times10^{-10}$. The local birth of a negative interval retains its nondegenerate fold structure at these points.

\begin{table}[tb]
\caption{Critical points for passage 27 from direct propagation with 20 ring modes, 24 translational states, and total occupation cutoff two. The coefficients $A$ and $B$ are the inverse-mass slope and phase curvature of the mixed response.}
\label{tab:epsilon-fold}
\setlength{\tabcolsep}{3pt}
\begin{ruledtabular}
\begin{tabular}{rrrrrr}
$\epsilon$ & $\mu_c$ & $M_c/M_0$ & $\delta_c$ & $10^7A$ & $10^4B$\\
0.00 & 1.03174 & 0.9692 & -0.24727 & -5.360 & 2.751\\
0.25 & 0.77764 & 1.2859 & -0.24530 & -5.339 & 2.639\\
0.50 & 0.53090 & 1.8836 & -0.24353 & -5.349 & 2.557\\
0.75 & 0.29260 & 3.4176 & -0.24207 & -5.388 & 2.502\\
1.00 & 0.06314 & 15.84 & -0.2410 & -5.453 & 2.471\\
\end{tabular}
\end{ruledtabular}
\end{table}

The inverse critical mass decreases from $1.03174$ to $0.06314$, giving a mass ratio of $16.34$ between the two endpoints. Over successive quarter-steps in $\epsilon$, the secant slopes of $\mu_c$ are $-1.0164$, $-0.9870$, $-0.9532$, and $-0.9178$. The large mass shift accompanies a smooth inverse-mass displacement with a finite response slope. Equation~\eqref{eq:mass-sensitivity} identifies its amplification by $1/\mu_c$. Using the final secant slope gives a local relative-sensitivity estimate of about $14.5$ per unit $\epsilon$. Figure~\ref{fig:epsilon-fold} displays both variables and the dependence on mode truncation.

\begin{figure*}[t]
\centering
\includegraphics[width=0.95\textwidth]{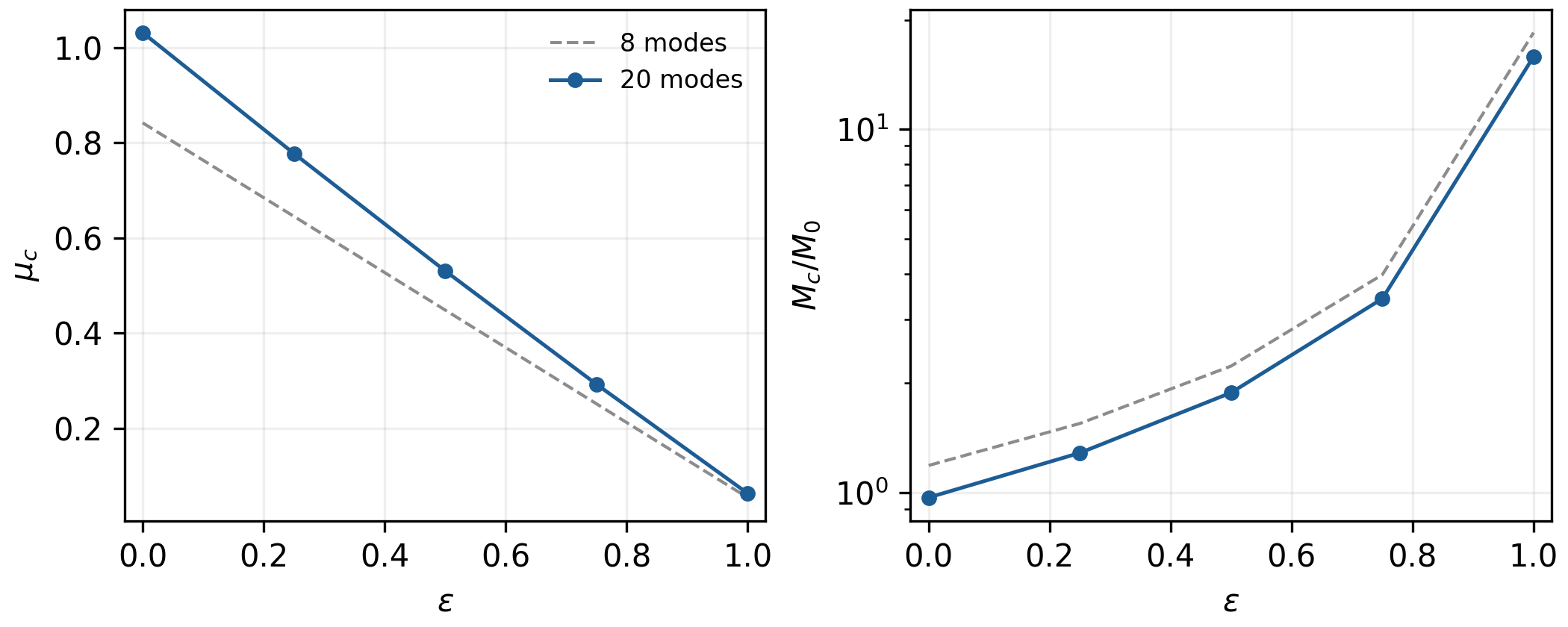}
\caption{Continuation of the passage-27 fold in cubic strength: inverse critical mass (left) and critical mass on a logarithmic scale (right). Solid curves connect the direct twenty-mode critical points in Table~\ref{tab:epsilon-fold}; dashed curves give the eight-mode continuation. Both calculations use 24 translational states and total occupation cutoff two.}
\label{fig:epsilon-fold}
\end{figure*}

The endpoint at $\epsilon=1$ is tested by independently increasing the number of ring modes from 20 to 24 and the translational basis from 24 to 32 states. At the original critical coordinates, these refinements give $F=-1.86\times10^{-12}$ and $F=-5.82\times10^{-12}$, respectively. Applying the local fold-displacement formula gives critical-mass estimates $15.8377M_0$ and $15.8395M_0$. These are Newton estimates from the refined response and phase derivatives. At $\epsilon=0$, increasing the mode count to 24 changes the response at the critical point by $7.22\times10^{-13}$.

A time-integration check at the cubic endpoint halves the maximum step from $0.08$ to $0.04$ and tightens the relative tolerance from $2\times10^{-12}$ to $2\times10^{-13}$, with the absolute tolerance set to one hundredth of the relative tolerance. The mixed response changes by $5.75\times10^{-14}$.

The occupation-cutoff refinement uses 20 ring modes and 24 translational states. Total occupation cutoffs two and three retain 231 and 1771 Fock states, respectively. The added sector admits $2\leftrightarrow3$ transitions from the linear mode vertices and $1\leftrightarrow3$ transitions from the pair part of $p:\mathcal W_2:$. Both the joint and the control evolutions use the same occupation cutoff.

At the two-quantum critical coordinates for passage 27, increasing the cutoff changes $\Gamma_{XR,2}$ by $5.00\times10^{-9}$ and $\Gamma_{R,2}$ by $4.47\times10^{-9}$, with $|\Delta\Gamma_{X,2}|<2\times10^{-15}$. Their difference gives $\Delta F=5.30\times10^{-10}$. With the same centered phase step $0.0005$ for both cutoffs, $\Delta F_\delta=-1.82\times10^{-8}$ and the phase curvature increases by $0.427\%$. The corresponding twelve-mode cutoff increment is $5.12\times10^{-10}$ in $F$.

Direct relocation of the three-quantum fold gives
\[
 \mu_c\simeq0.06411,\qquad \delta_c\simeq-0.24092,
 \qquad M_c\simeq15.60M_0.
\]
At the computed coordinates, $F=-4.48\times10^{-14}$ and $F_\delta=8.10\times10^{-10}$. The linearized mass correction inferred from the residual is $2.0\times10^{-5}M_0$. The derivatives $A=-5.450\times10^{-7}$ and $B=2.481\times10^{-4}$ retain the nondegenerate birth of a negative interval, with opening coefficient $s=0.06628$. Relative to the two-quantum result, the critical mass decreases by $0.2392M_0$, or $1.51\%$.

Table~\ref{tab:cutoff20-convergence} compares the independent numerical refinements. The occupation-cutoff change produces the largest mass displacement among these tests. The three-quantum calculation supplies the refined endpoint at $\epsilon=1$, while Table~\ref{tab:epsilon-fold} specifies the continuation at occupation cutoff two.

\begin{table}[tb]
\caption{Critical-mass estimates for the center--mode family at passage 27 and $\epsilon=1$. Packet denotes the number of translational states; cutoff bounds the total mode occupation. Changes are relative to the 20-mode, 24-state, two-quantum reference. Newton estimates use the local fold-displacement formula; direct entries satisfy the two fold equations within the stated residual tolerances.}
\label{tab:cutoff20-convergence}
\setlength{\tabcolsep}{2pt}
\begin{ruledtabular}
\begin{tabular}{rrrrrl}
Modes & Packet & Cutoff & $M_c/M_0$ & $\Delta M_c/M_0$ & Method\\
20 & 24 & 2 & 15.84 & 0 & Direct\\
24 & 24 & 2 & 15.84 & $+0.0009$ & Newton\\
20 & 32 & 2 & 15.84 & $+0.0027$ & Newton\\
20 & 24 & 3 & 15.60 & $-0.24$ & Direct\\
\end{tabular}
\end{ruledtabular}
\end{table}

Spatial-grid refinement from 192 to 256 points compares the first 24 modes after aligning their complex phases. The maximum changes are $8.74\times10^{-14}$ in mode frequency, $1.36\times10^{-13}$ in a source projection, and $3.3\times10^{-16}$ in a cubic coefficient. The corresponding maximum frequency-phase change over 27 passages is $5.25\times10^{-11}$.

% BEGIN FULL FIELD RESPONSE AUDIT
\subsection{Field-order completion at fixed mode cutoff}
\label{sec:full-field-response}
We propagate the twenty-mode field reduction through fourth degree,
\begin{equation}
\begin{split}
 \mathcal H_{\rm f4}={}&\mathcal H_2(\mu,\delta)+\hstar H_q
                 +\widetilde{\mathcal H}_3+\mathcal H_4\\
 &-\kappa Z_1\operatorname{Op}_{\rm W}(S_0+S_1+S_2).
\end{split}
\label{eq:full-field-four}
\end{equation}
Here $\widetilde{\mathcal H}_3$ uses the fixed-reference-velocity subtraction of Appendix~\ref{app:canonical-vertices}; $\mathcal H_4$ and $S_n$ are given by Eqs.~\eqref{eq:quartic-field} and \eqref{eq:source-hierarchy}. All field coefficients retain their reference values as $\mu$ and the diagnostic detuning vary. Direct propagation of Eq.~\eqref{eq:full-field-four} includes repeated cubic interactions and their interference with the source. The joint evolution and both controls receive the corresponding terms from the same Hamiltonian.
Put $X_Q=vt-L/2+Q$, and let $c_4$ be the coefficient of $p^4$ in the canonical fourth-degree energy. The translational control used here is
\[
\begin{split}
 \mathcal H_X={}&\frac{\mu p^2}{2M_0}+c_3p^3+c_4p^4+\hstar H_q\\
 &-\kappa Z_1\left[J(X_Q)+\tfrac12\{g_\alpha(X_Q),p\}\right].
\end{split}
\]
It keeps the pure-momentum polynomial and the momentum source, with the oscillator coordinates set to zero in the canonical symbols. The radiative control is the joint Hamiltonian evaluated at $Q=p=0$, acting on the chain and the retained modes. In the matched comparison it uses the same frequency counterterms and the same initial mode state as the joint evolution. The translational control has quadratic curvature $\mu/M_0$ and retains the displayed $p^3$ and $p^4$ terms. The mean-source control below replaces the source in this same translational Hamiltonian by its expectation in the matched mode ground state.

The comparison is made at the three-quantum center--mode endpoint,
\[
 N=27,\qquad \mu=0.06411253,\qquad\delta=-0.24092163.
\]
Weyl polynomials are normally ordered before projection on the oscillator basis, retaining their vacuum contractions. The kinetic ordering potential inherited from the coordinate measure has the form $\hstar^2V_{\rm ord}(q)$. Its fluctuation-independent part contributes a common phase at fourth degree, and its first variable term has energy order $\hstar^{5/2}$ under the counting in Appendix~\ref{app:field-reduction}.

For propagation, we use 24 translational states and oscillator states satisfying both $\sum_\nu n_\nu\leq4$ and $\sum_\nu\omega_\nu^{\rm ref}n_\nu\leq E_{\max}$. The latter bound uses the unshifted reference frequencies, keeping the basis fixed during phase differentiation. The twenty-mode vacuum contractions are evaluated before this projection. All twenty one-quantum states are present at the tabulated energy bounds. DOP853 integration uses relative tolerance $10^{-9}$, absolute tolerance $10^{-11}$, and maximum step $0.4$.

Table~\ref{tab:full-field-blocks} resolves the contribution of the additional blocks in the same basis, with $E_{\max}=4$. Completing the cubic energy and momentum source changes $F$ by $-4.6586\times10^{-7}$. Adding the fourth-degree energy and quadratic source gives a compensating change of $+1.0297\times10^{-7}$. The resulting point has $F<0$ and hence belongs to an open suppression region. Its total conditional loss also decreases from $5.61007\times10^{-5}$ to $4.99069\times10^{-5}$.

\begin{table}[t]
\caption{Field-block comparison at passage 27 in a common 231-state oscillator basis ($E_{\max}=4$). The middle row adds the pure-mode cubic block and the momentum source to the center--mode model. The final row additionally includes $\mathcal H_4$ and $S_2$.}
\label{tab:full-field-blocks}
\setlength{\tabcolsep}{3pt}
\begin{ruledtabular}
\begin{tabular}{lrrrr}
Model & $10^5\Gamma_{XR,2}$ & $10^5\Gamma_{R,2}$ & $10^5\Gamma_{X,2}$ & $10^7F$\\
Center--mode & 5.61007 & 1.37405 & 4.23620 & $-0.01748$\\
Complete cubic & 5.98630 & 1.84202 & 4.19104 & $-4.67608$\\
Fourth degree & 4.99069 & 0.83611 & 4.19104 & $-3.64636$
\end{tabular}
\end{ruledtabular}
\end{table}

Table~\ref{tab:full-field-energy} reports the energy-cutoff check. We also compute the spectrum of the source-free mode block. Its lowest excitation, with squared overlap $0.9876$ with the first bare-mode state, has a frequency increment of approximately $1.17\times10^{-3}$. The corresponding passage-phase increment approaches $0.02598$ as $E_{\max}$ increases to $5.5$, giving an accumulated phase scale of $0.70$ at passage 27. The fourth-degree vacuum coefficient of $p^2$ also gives an inverse-curvature increment $2M_0\langle0|\mathcal H_4^{(p^2)}|0\rangle=0.00279754$. These spectral and kinetic contributions enter together in the propagated response.

Tightening the relative tolerance to $10^{-11}$ and reducing the maximum step to $0.2$ changes $F$ at $E_{\max}=4$ by $4.49\times10^{-12}$. At that energy bound, centered phase differences with step $0.005$ give $F_\delta\simeq4.61\times10^{-6}$ and $F_{\delta\delta}\simeq1.25\times10^{-4}$; a forward inverse-mass step of $0.02$ gives $F_\mu\simeq-4.36\times10^{-7}$. At the boundary $\mu=0$ of the diagnostic curvature family, the same phase gives $F=-3.36\times10^{-7}$. The negative response is also present at the two neighboring phases and at $\mu=0.08411$.

The normal-order contractions additionally depend on the total number of modes. A separate source-free comparison retains all states through three quanta at each mode count, with the same bare reference coefficients. The first excitation has passage-phase increments $0.02598$, $0.03645$, and $0.06467$ for 20, 24, and 32 modes, respectively. Energy-cutoff refinement and total-mode refinement change different parts of the regulated Hamiltonian. Tables~\ref{tab:full-field-blocks} and \ref{tab:full-field-energy} specify the twenty-mode response at fixed reference coefficients.

\begin{table}[t]
\caption{Energy-cutoff check for the fourth-degree Hamiltonian at the former center--mode fold. The occupation bound is four and the packet basis contains 24 states. The phase increment is obtained from the lowest excitation of the source-free mode block.}
\label{tab:full-field-energy}
\begin{ruledtabular}
\begin{tabular}{rrrr}
$E_{\max}$ & Oscillator states & $10^7F$ & $\Delta\omega_1T$\\
4 & 231 & $-3.64636$ & 0.0283402\\
4.5 & 403 & $-3.90355$ & 0.0260648\\
5 & 665 & $-3.88949$ & 0.0260122\\
\end{tabular}
\end{ruledtabular}
\end{table}
% END FULL FIELD RESPONSE AUDIT

\subsection{Comparison at fixed excitation gaps and translational mass}
\label{sec:matched-response}
For each mode count we specify the source-free one-particle excitation gaps and the curvature of the ground-branch energy as physical parameters of the finite-mode model. Appendix~\ref{app:physical-matching} gives the counterterms that impose these conditions. The physical passage phase is $\delta_{\rm phys}=\omega_{1,\rm phys}T-6\pi=-0.24092163$, and the reference inverse mass is $\mu=0.06411253$. The source geometry, nonlinear vertices, chain Hamiltonian, passage period, and initial Gaussian packet are common to all calculations. We compare two factorized initial mode states: the harmonic vacuum and the matched ground state at $p=0$.

% BEGIN MATCHED TABLE RESULTS
Table~\ref{tab:matched-response} compares the mixed response at $N=27$. The negative sign persists at 16, 20, and 24 modes for both initial preparations. Since the finite-dimensional propagator depends continuously on phase and mass, each strictly negative value lies in an open suppression region. At fixed 20 modes, raising the energy bound from $4.5$ to $5$ changes $F$ by $3.75\times10^{-9}$ for the harmonic vacuum and $6.82\times10^{-10}$ for the matched ground state. Changing the mode count from 20 to 24 at energy bound 5 changes the respective magnitudes by approximately $37\%$. Across these mode counts, the sign is preserved and the magnitude changes.

\begin{table}[t]
\caption{Matched-parameter mixed response at passage 27. Both preparations use 24 packet states and total mode occupation at most four. The last two columns correspond to the harmonic vacuum and the matched ground state at zero momentum. The physical phase and mass are fixed at $\delta_{\rm phys}=-0.24092163$ and $M/M_0=15.5976$.}
\label{tab:matched-response}
\begin{ruledtabular}
\begin{tabular}{rrrr}
Modes & $E_{\max}$ & $10^7F_{|0\rangle}$ & $10^7F_{|g_K\rangle}$\\
16 & 5 & $-3.15270$ & $-5.16148$\\
20 & 4.5 & $-4.73956$ & $-7.71349$\\
20 & 5 & $-4.70206$ & $-7.70667$\\
24 & 5 & $-6.45096$ & $-10.57414$\\
\end{tabular}
\end{ruledtabular}
\end{table}

Tightening the relative and absolute integration tolerances to $10^{-11}$ and $10^{-13}$ and reducing the maximum step from $0.4$ to $0.2$ changes the 20-mode, $E_{\max}=4.5$ harmonic-vacuum result by $4.439\times10^{-12}$.

At the same physical phase and $\mu=0.1$ ($M=10M_0$), the harmonic-vacuum response is $F=-4.96104\times10^{-7}$ for 20 modes and $F=-6.73584\times10^{-7}$ for 24 modes.
% END MATCHED TABLE RESULTS

The translational control in Eq.~\eqref{eq:mixed} evaluates the source at zero mode coordinates. For the matched ground-state preparation, we also form a separate control by taking the expectation of the source operator in $|g_K\rangle$:
\[
 \overline{S}_K(Q,p)=\langle g_K|\operatorname{Op}_{\rm W}(S_0+S_1+S_2)|g_K\rangle.
\]
This control retains the same packet dispersion and chain Hamiltonian. Denoting its loss by $\overline{\Gamma}_{X,2}$, define
\[
 \overline F=\Gamma_{XR,2}-\Gamma_{R,2}-\overline{\Gamma}_{X,2},
 \qquad F=\overline F+\overline{\Gamma}_{X,2}-\Gamma_{X,2}.
\]
At the reference mass and energy bound 5, $\overline{\Gamma}_{X,2}=4.11847\times10^{-5}$ for 20 modes and $4.09159\times10^{-5}$ for 24 modes. The resulting residuals are $\overline F=-4.50\times10^{-8}$ and $-6.30\times10^{-8}$, respectively. The average source profile accounts for most of the negative ground-state response, while the residual is also negative at both mode counts. The Fourier representation of the averaged quadratic source agrees with direct ground-state matrix elements to below $10^{-18}$.

\subsection{Excitation probabilities and conditional spectral weights}
\label{sec:spectral-weights}
We resolve the matched-ground-state calculations of Appendix~\ref{sec:matched-response} in the eigenbasis of the source-free mode Hamiltonian at zero momentum. Let $K$ denote the number of retained modes, $|g_K\rangle$ its ground state, and $|r;K\rangle$ a complete orthonormal eigenbasis in the retained oscillator space. The one-particle branch $r_\nu$ is identified by its overlap with the reference harmonic state $|1_\nu\rangle$, using the same assignment as in Appendix~\ref{app:physical-matching}. For a normalized joint state, define
\begin{equation}
 A_{rj}(Q)=\langle r;K|\Psi_j(Q)\rangle,\qquad
 P_r=\sum_j\int|A_{rj}(Q)|^2\dd Q .
 \label{eq:branch-probability}
\end{equation}
Here $j=0,1,2,3$ labels the vacuum and one-excitation states of the chain. The initial factorized preparation has $P_g=1$. Equation~\eqref{eq:branch-probability} measures excitation of the matched eigenbranches during the joint evolution.

Put $p_j=\rho_{jj}$ and $\vartheta=\arg\rho_{20}$. The corresponding conditional spectral weights are
\begin{equation}
\begin{gathered}
 D_r=\frac12\int\left|
 \frac{A_{r2}(Q)}{\sqrt{p_2}}
 -e^{i\vartheta}\frac{A_{r0}(Q)}{\sqrt{p_0}}
 \right|^2\dd Q,\\
 D_g+\sum_{\nu=1}^K D_{r_\nu}+D_{\rm rest}=1-\eta_2 .
\end{gathered}
\label{eq:branch-distance}
\end{equation}
The remainder sums the other eigenstates. Orthogonality gives Eq.~\eqref{eq:branch-distance} directly by resolving the squared distance between the two normalized conditional environmental vectors. Its ground-state term retains the dependence on the translational coordinate $Q$.

\begin{figure*}[t]
\centering
\includegraphics[width=\textwidth]{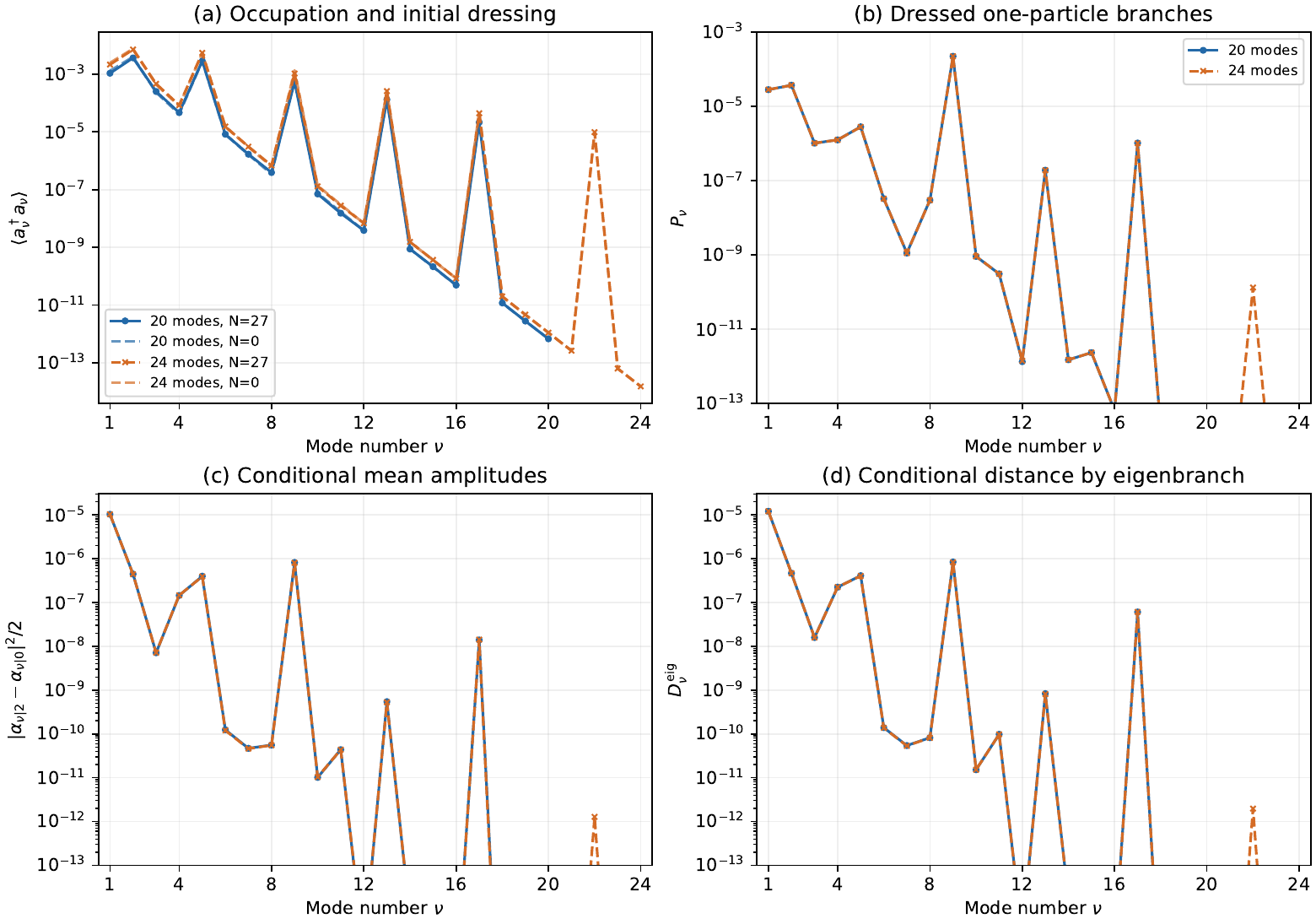}
\caption{Spectral diagnostics at passage 27 for 20 and 24 modes, with matched excitation gaps and translational curvature. (a) Reference harmonic occupations at preparation and after propagation. (b) One-particle eigenbranch probabilities $P_{r_\nu}$ from Eq.~\eqref{eq:branch-probability}. (c) Differences of conditional harmonic means. (d) Eigenbranch contributions $D_{r_\nu}$ to Eq.~\eqref{eq:branch-distance}. The eigenbranch labels follow their harmonic one-quantum overlaps. Panels (b)--(d) display values above $10^{-13}$.}
\label{fig:mode-distribution}
\end{figure*}

For comparison with the reference harmonic basis, we also record $n_\nu=\langle a_\nu^\dagger a_\nu\rangle$ and the conditional means
\[
 \alpha_{\nu|j}=p_j^{-1}\langle\Psi_j|a_\nu|\Psi_j\rangle.
\]
Figure~\ref{fig:mode-distribution} displays these moments together with the eigenbranch probabilities and distances at passage 27. The physical phase, mass, source, and initial packet are those of Table~\ref{tab:matched-response}, with $E_{\max}=5$ and the matched ground-state preparation.

The ninth eigenbranch carries $75.65\%$ and $75.68\%$ of the summed one-particle excitation probability for 20 and 24 modes, respectively. The first eigenbranch carries $85.66\%$ and $85.83\%$ of the summed one-particle conditional distance. The two observables select different dominant branches. This dependence on the measured quantity also occurs in reductions of logical channels~\cite{gordeychuk2026codes} and quantum phase estimation~\cite{gordeichuk2026qpe}. In the present calculation, Eqs.~\eqref{eq:branch-probability} and \eqref{eq:branch-distance} give the weights explicitly from the joint state: excitation measures the occupied branch norm, while conditional distance measures the separation of the two normalized conditional amplitudes.

Table~\ref{tab:spectral-weights} quantifies the mode-count comparison. The summed one-particle probability changes by $0.118\%$ between 20 and 24 modes, while the summed one-particle distance changes by $-0.0258\%$. The added branches 21--24 have total probability $1.32\times10^{-10}$ and distance $1.97\times10^{-12}$. The ground-state projection accounts for $98.18\%$ of the change in the full conditional distance. Equation~\eqref{eq:branch-distance} assigns this change to the ground-state and excited sectors.

\begin{table}[t]
\caption{Eigenbranch diagnostics at passage 27 for the matched-ground-state preparation. The sums run over the identified one-particle branches. $D_{\rm rest}$ contains all other excited eigenstates.}
\label{tab:spectral-weights}
\begin{ruledtabular}
\begin{tabular}{lrr}
Quantity & 20 modes & 24 modes\\
$10^4\sum_\nu P_{r_\nu}$ & 2.90538 & 2.90880\\
$10^5\sum_\nu D_{r_\nu}$ & 1.40204 & 1.40168\\
$10^5D_g$ & 4.08511 & 4.05821\\
$10^8D_{\rm rest}$ & 4.66718 & 4.53081\\
$10^5(1-\eta_2)$ & 5.49181 & 5.46442\\
\end{tabular}
\end{ruledtabular}
\end{table}

The harmonic ground-state occupations provide a separate measure of the background deformation. Their sums are $0.00919027$ and $0.01757328$ for 20 and 24 modes, whereas the added harmonic modes contribute $9.77\times10^{-6}$. More than $99.99\%$ of each initial occupation sum is accounted for by the squared mean amplitudes. Appendix~\ref{app:mode-diagnostics} relates this displacement to the cubic vacuum contractions and reports the passage-resolved occupation changes. The summed one-particle excitation and distance weights change by less than $0.12\%$ between these two cutoffs. The mean-source control in Appendix~\ref{sec:matched-response} measures the accompanying change in the average interaction profile.

\section{Sector-resolved conditional distance}
Let $E_j^{(n)}$ denote the environmental vector conditional on chain state $j$ and total oscillator occupation $n$, including its dependence on $\xi$. Put $p_j=\rho_{jj}$ and $\vartheta_j=\arg\rho_{j0}$. Direct expansion of the squared norm gives
\[
 1-\eta_j=\frac12\sum_n\left\|
 \frac{E_j^{(n)}}{\sqrt{p_j}}
       -e^{i\vartheta_j}\frac{E_0^{(n)}}{\sqrt{p_0}}\right\|^2.
\]
This identity expresses the loss as a sum of nonnegative sector contributions. At $N=16$ the zero-, one-, and two-quantum contributions for qubit 2 are $0.7946\times10^{-6}$, $1.5880\times10^{-6}$, and $0.0060\times10^{-6}$. At $N=31$ they are $2.1211\times10^{-6}$, $1.8902\times10^{-6}$, and $0.0073\times10^{-6}$. These values follow from the joint 20-mode, two-quantum calculation. The vacuum sector includes the translated packet and returns from excited sectors.

\section{A kinematic scale for the loss of passage memory}
\label{app:kinematic-memory}
For an isolated spatial Fourier vertex $e^{iq\xi}$ and free translational kinetic energy, recoil adds the phase
\[
 \Delta\phi(p)=\frac{T[(p+\hstar q)^2-p^2]}{2M\hstar}
             =\frac{Tqp}{M}+\frac{\hstar Tq^2}{2M}.
\]
In this appendix $\delta$ is the detuning phase of the selected scalar Fourier channel, reduced modulo $2\pi$. A Gaussian momentum distribution of width $\hstar/(2\sigma_X)$ gives mean recoil $\delta_{\rm rec}=\hstar Tq^2/(2M)$ and phase spread $s_\phi=\hstar T|q|/(2M\sigma_X)$. Averaging the scalar filter produces
\[
 \overline G_N=N+2\sum_{\ell=1}^{N-1}(N-\ell)
 e^{-\ell^2s_\phi^2/2}\cos[\ell(\delta+\delta_{\rm rec})].
\]
The Gaussian factors $e^{-\ell^2s_\phi^2/2}$ are the characteristic function of the packet-induced phase spread at harmonic $\ell$. They weight the Fej\'er sum by the surviving correlation between passages separated by $\ell$ turns. In the coherence-transfer representation of QPE, the harmonic weights are set by the phase channel~\cite{gordeichuk2026qpe}. Here they are determined by the momentum distribution of the fluxon packet.

For $z=N\delta$, $\rho=N\delta_{\rm rec}$, and $\beta=Ns_\phi$,
\[
 \frac{\overline G_N}{N^2}\longrightarrow
 2\int_0^1(1-u)e^{-\beta^2u^2/2}\cos[(z+\rho)u]\dd u.
\]
The expansion at small $\rho,\beta$ is $f(z)+\rho f'(z)+(\rho^2+\beta^2)f''(z)/2$. This gives a kinematic memory scale $N\sim1/s_\phi$. Source-weighted root-mean-square wave numbers for modes 1, 5, and 9 give estimates $36s_\phi\simeq0.078$, $0.049$, and $0.092$ at $M_0$. The full sign boundaries above are obtained from the operator response, including source-weight changes and vacuum returns.

\section{Field reduction and the order of translation--mode couplings}
\label{app:field-reduction}
\subsection{Constrained coordinates and the kinetic metric}
Use dimensionless energy units $E_0$ in this appendix; division of a Hamiltonian by $\hstar$ gives the frequency units of Eq.~\eqref{eq:hamiltonian}. With a normalized chain vector $c$, the action is
\begin{equation}
\begin{split}
 \mathcal L={}&\int_0^L\left[\frac{\phi_t^2}{2}
 -\frac{\phi_x^2}{2}-(1-\cos\phi)\right]\dd x
 +i\hstar c^\dagger\dot c-\hstar c^\dagger H_qc\\
 &-\kappa(c^\dagger Z_1c)\int_0^LG(x)\phi_x\dd x .
\end{split}
\label{eq:field-action}
\end{equation}
Equation~\eqref{eq:field-action} takes the effective chain Hamiltonian as input. Its field part determines the collective kinetic energy and the source projections.
Choose real periodic functions $e_a$ satisfying
\[
 \langle e_a,e_b\rangle=\delta_{ab},\qquad
 \langle e_a,\Phi_v'\rangle=0,
 \qquad \langle f,g\rangle=\int_0^Lfg\dd y.
\]
At fixed reference velocity $v$, introduce the local coordinate chart
\begin{equation}
\begin{gathered}
 \phi(x,t)=\Phi_v(y)+\eta(y,t),\\
 y=x-X(t),\qquad \eta=\sum_a q_a(t)e_a(y).
\end{gathered}
 \label{eq:field-chart}
\end{equation}
In this appendix $q=(q_a)$ denotes real deformation coordinates and $\pi=(\pi_a)$ their conjugate momenta; primes denote derivatives in $y$. The orthogonality in Eq.~\eqref{eq:field-chart} fixes the center locally. The real coordinate basis is converted to the moving normal modes by a symplectic diagonalization of the quadratic dynamics. Substitution gives $\phi_t=\sum_a\dot q_ae_a-\dot X(\Phi_v'+\eta')$. Define
\[
\begin{gathered}
 I_0=\langle\Phi_v',\Phi_v'\rangle,\quad
 f_a=\langle\Phi_v',e_a'\rangle,\\
 D_{ab}=\langle e_a,e_b'\rangle=-D_{ba},\\
 C_{ab}=\langle e_a',e_b'\rangle,\quad
 b_a=\sum_bD_{ab}q_b,\\
 I(q)=I_0+2f^Tq+q^TCq.
\end{gathered}
\]
These integrals define the scalar $I_0$, vector $f$, and matrices $C,D$, with $b=Dq$. In this appendix $T$ denotes kinetic energy. The kinetic energy and its Schur complement are
\begin{equation}
 T=\frac12\dot q^T\dot q-\dot X b^T\dot q
                  +\frac12 I(q)\dot X^2,\qquad
 m(q)=I(q)-b^Tb.
 \label{eq:kinetic-metric}
\end{equation}
For a complete transverse basis, completeness gives
$C-D^TD=ff^T/I_0$, so $m(q)=(I_0+f^Tq)^2/I_0$ locally. At a finite mode cutoff the definitions in Eq.~\eqref{eq:kinetic-metric} give the Galerkin metric directly.

\subsection{Legendre transform and moving-frame expansion}
The momenta are $\pi=\dot q-\dot Xb$ and
$P=I\dot X-b^T\dot q=m\dot X-b^T\pi$.
Writing $B=b^T\pi$, the classical collective Hamiltonian is
\begin{equation}
 \mathcal H=\frac12\pi^T\pi+\frac{(P+B)^2}{2m(q)}
       +V(q)+\hstar H_q+\mathcal V_{\rm int}(X,q),
 \label{eq:metric-hamiltonian}
\end{equation}
where
\[
\begin{split}
 V(q)&=\int_0^L\left[\frac12(\Phi_v'+\eta')^2
                           +1-\cos(\Phi_v+\eta)\right]\dd y,\\
 \mathcal V_{\rm int}&=-\kappa Z_1
       \left[J(X)+\sum_aq_a g_a(X)\right],\\
 g_a(X)&=\int_0^Lg(y+X)e_a(y)\dd y.
\end{split}
\]
Integration by parts yields this source term and Eq.~\eqref{eq:projections}. The source is exactly linear in the field deformation in this coordinate chart.

For $X=vt-L/2+\xi$, $P=vI_0+p$, and $d(q)=m(q)-I_0$, the kinetic part of $\mathcal H-vP$, up to a constant, is exactly
\begin{equation}
 \frac12\pi^T\pi+vB-\frac{v^2d}{2}
       +\frac{[p+B-vd]^2}{2[I_0+d]}.
 \label{eq:moving-metric}
\end{equation}
Equation~\eqref{eq:moving-metric} specifies all kinetic vertices before diagonalization. For example, its quadratic terms contain
$-2vp f^Tq/I_0$; its higher terms include $pB/I_0$ and
$-p^2f^Tq/I_0^2$, together with the remaining terms from the same square and denominator. Their coefficients must be transformed together. The field potential additionally contains
\[
 V_3=-\frac16\int_0^L\sin\Phi_v\,\eta^3\dd y,
 \qquad
 V_4=-\frac1{24}\int_0^L\cos\Phi_v\,\eta^4\dd y.
\]

The velocity dependence of the profile accounts for part of the kinetic coupling already at quadratic order. Write $K$ for the transverse stiffness matrix in this appendix. Define it by
\[
 K_{ab}=\left\langle e_a,
 [-(1-v^2)\partial_y^2+\cos\Phi_v]e_b\right\rangle.
\]
Differentiate the profile equation at fixed circumference, choosing the center so that $\chi=\partial_v\Phi_v$ is transverse. Then $K\chi=2vf$ in transverse coordinates. Consequently,
\begin{equation}
 \frac{\dd P}{\dd v}=I_0+4v^2 f^TK^{-1}f=M_0.
 \label{eq:dressed-mass}
\end{equation}
The inverse in Eq.~\eqref{eq:dressed-mass} acts on the transverse subspace. At $L=12$ and $v=0.54$, the rigid-profile metric is $I_0=9.50534199$, while Eq.~\eqref{eq:dressed-mass} gives $M_0=13.41599870$. Fourier grids of 96, 192, and 384 points agree to $2\times10^{-11}$ in the latter value; a centered finite difference of $P(v)$ with step $10^{-5}$ agrees to $4\times10^{-9}$. This relation gives the profile-relaxation contribution to Eq.~\eqref{eq:mass}. The quadratic reduction combines this relaxation with the gyroscopic term $vB$ to produce the moving normal-mode problem. The diagonalized quadratic Hamiltonian includes the profile relaxation generated by the $pq$ vertex. The same canonical transformation acts on the source and on the physical center observable.

At a regulated finite-dimensional level, quantization in these coordinates uses the field-induced measure $\sqrt{\det\mathsf g}\,\dd X\,\dd q$, where
\[
 \mathsf g=\begin{pmatrix}I&-b^T\\-b&\mathbf1\end{pmatrix},
 \qquad \det\mathsf g=m.
\]
The coordinate kinetic operator inherited from the flat field kinetic energy is
\[
 \widehat T=-\frac{\hstar^2}{2\sqrt{\det\mathsf g}}
 \partial_A\left(\sqrt{\det\mathsf g}\,
                    \mathsf g^{AB}\partial_B\right).
\]
Quantization of Eq.~\eqref{eq:metric-hamiltonian} fixes both the measure and the operator ordering through the same field regulator.

\subsection{Field order and accumulation over many passages}
\label{sec:field-order}
The parameter $\lambda$ in Eq.~\eqref{eq:hamiltonian} counts source vertices that create or remove oscillator quanta. The metric interactions in Eq.~\eqref{eq:metric-hamiltonian} carry a separate expansion in field fluctuations. An insertion of a geometric vertex into a two-source response contributes at the same source order $\lambda^2$ as Eq.~\eqref{eq:mixed-order}.

The full cubic energy is given in Eq.~\eqref{eq:cubic-vertices}. Put $s=f^Tq$, $d_2=q^T(C-D^TD)q$, $r=p-2vs$, and $t_2=B-vd_2$. Expansion of Eq.~\eqref{eq:moving-metric} gives the homogeneous fourth-degree energy
\begin{equation}
\begin{split}
 \mathcal H_4={}&\frac{t_2^2}{2I_0}-\frac{2srt_2}{I_0^2}
 +\left(\frac{2s^2}{I_0^3}-\frac{d_2}{2I_0^2}\right)r^2\\
 &-\frac1{24}\int_0^L\cos\Phi_v\,\eta^4\dd y.
\end{split}
\label{eq:quartic-field}
\end{equation}
The same canonical map~\eqref{eq:canonical-map} acts on all terms in Eq.~\eqref{eq:quartic-field}. Comparing its kinetic polynomial with the exact rational kinetic energy along three independent phase-space directions gives a fifth-degree remainder: halving the amplitude reduces the error by approximately $32$. Quantization inherits the field kinetic operator and its measure specified above; its ordering terms must be carried with the fourth-degree energy. The scalar energy offset cancels from conditional overlaps.

The source expansion also follows from that map. Write $\xi=Q+\zeta$, where $Q$ is the canonical center coordinate and $\zeta$ its oscillator-dependent displacement in Eq.~\eqref{eq:canonical-map}. Put $q=\alpha p+\widetilde q$, with $\widetilde q$ the oscillator part of the deformation, and $g_a(X)=\int_0^Lg(y+X)e_a(y)\dd y$. With all functions evaluated at $X_Q=vt-L/2+Q$, its Weyl symbol is $-\kappa Z_1\sum_{n\geq0}S_n$, where
\begin{equation}
\begin{aligned}
 S_0&=J,\qquad S_1=J'\zeta+g^T(\alpha p+\widetilde q),\\
 S_n&=\frac{J^{(n)}}{n!}\zeta^n
 +\frac{\zeta^{n-1}}{(n-1)!}(g^{(n-1)})^T
                         (\alpha p+\widetilde q),\quad n\geq2.
\end{aligned}
\label{eq:source-hierarchy}
\end{equation}
Here $Q$ is kept exact. The term $S_2$ contains both $J''\zeta^2/2$ and $\zeta(g')^T(\alpha p+\widetilde q)$. The linear term in Eq.~\eqref{eq:source-hierarchy} gives the oscillator source and the momentum source in Eq.~\eqref{eq:source-momentum}.

For $q,\pi,p=O(\sqrt{\hstar})$ and $\kappa/\hstar=O(1)$, the frequency Hamiltonian has an $O(\sqrt{\hstar})$ block
\[
 A_1=(\mathcal H_3-\kappa Z_1S_1)/\hstar
\]
and an $O(\hstar)$ block $A_2=(\mathcal H_4-\kappa Z_1S_2)/\hstar$, with the inherited ordering contributions included. The leading propagator contains the quadratic energy, the chain Hamiltonian, and $-\kappa Z_1S_0$. Its interaction-picture expansion through frequency order $\hstar$ contains
\[
 -i\int A_{1,I}\dd t-i\int A_{2,I}\dd t
 -\int_{t_1>t_2}A_{1,I}(t_1)A_{1,I}(t_2)\dd t_1\dd t_2.
\]
This order includes the full cubic interaction, the quartic interaction, the first two transformed source terms, and ordered pairs of cubic and linear-source insertions. A normal-form construction transfers nonresonant cubic terms into fourth-degree coefficients and transforms the source and endpoint states at the same time. Collective-coordinate reductions with dynamical recoil are also developed in Ref.~\cite{melnikov2020}.

The packet used here has $\sigma_p=0.001491=0.06105\sqrt{\hstar}$. Under a uniform-in-time bound $p=O(\hstar)$, the leading center--mode term $p\mathcal W_2/\hstar$ is $O(\hstar)$; the pure-mode cubic term remains $O(\sqrt{\hstar})$, and its second insertion and the pure-mode quartic term are $O(\hstar)$. This recoil counting retains those mode terms together with the leading center--mode term. The initial packet width supplies the initial condition for the required momentum-moment bound.

Accumulation is determined by weighted insertion sums. Let $\mathcal V(t)$ be an energy vertex, $U_0(t,0)$ the retained coupled propagator, and $U_T=U_0(T,0)$. For a periodic retained Hamiltonian and vertex, define
\begin{equation}
\begin{aligned}
 K_{\mathcal V}&=\frac1{\hstar}\int_0^T
 U_0(t,0)^\dagger\mathcal V(t)U_0(t,0)\dd t,\\
 K_{\mathcal V}^{(N)}&=\sum_{n=0}^{N-1}
 (U_T^\dagger)^nK_{\mathcal V}U_T^n.
\end{aligned}
\label{eq:field-insertion-sum}
\end{equation}
If $U_T|a\rangle=e^{-i\theta_a}|a\rangle$, a matrix element of Eq.~\eqref{eq:field-insertion-sum} contains the factor
\[
 \sum_{n=0}^{N-1}e^{in\Delta\theta}
 =e^{i(N-1)\Delta\theta/2}
 \frac{\sin(N\Delta\theta/2)}{\sin(\Delta\theta/2)},
 \qquad\Delta\theta=\theta_a-\theta_b.
\]
Its magnitude is bounded by $\min(N,|\sin(\Delta\theta/2)|^{-1})$, with limiting value $N$ at resonance. The one-period matrix element fixes the weight. For a static vertex between source-free levels with frequency difference $\Omega$, the full weight is $(e^{i\Omega NT}-1)/(i\Omega)$. At $\Omega T=2\pi m$ with nonzero integer $m$, the one-period integral vanishes. The resonance criterion uses the product of the insertion matrix element and the passage sum. Resonant contributions of appreciable accumulated size belong in the propagated generator. For an energy remainder $R$ and the retained state $\psi_{\rm ret}$, Duhamel's formula gives the state-error bound
\[
 \|\psi(NT)-\psi_{\rm ret}(NT)\|
 \leq\frac1{\hstar}\int_0^{NT}
       \|R(t)\psi_{\rm ret}(t)\|\dd t.
\]
Uniform fluctuation-moment bounds and bounded coefficients give a nominal remainder scale $NT\hstar^{3/2}$ after the fourth-degree energy and quadratic source have been retained. For a specified state and coherence observable, we evaluate the weighted sums and the tangent response.

At passages 27 and 36, respectively, $NT\sqrt{\hstar}=14.65,19.53$ and $NT\hstar=0.358,0.477$. These are power-counting factors; the spectral weights determine the actual accumulated amplitudes. As a direct diagnostic, the Weyl-ordered pure-mode cubic block $\mathcal H_{3,\rm rad}=\mathcal H_3|_{p=0}$ was projected on twenty modes, retaining its linear Wick contractions. With source-free frequencies, its first-order action on the bare vacuum has norm $0.1156$, $0.0950$, and $0.1529$ at passages 18, 27, and 36. The corresponding operator norms from the vacuum-plus-one-quantum input space to sectors through four quanta are $0.2006$, $0.1654$, and $0.2666$. These quantities measure accumulated state amplitudes in the source-free diagnostic. The coupled-chain correction is obtained from Eq.~\eqref{eq:vertex-tangent}, with the same vertex included in each control defining the mixed response.

For a fold, this observable-level check is particularly restrictive. If $A=F_\mu$, $B=F_{\delta\delta}$, and $C=F_{\mu\delta}$ at the critical point, a small additional contribution $\Delta F$ displaces it by
\[
\begin{aligned}
 \Delta\mu_c&=-\frac{\Delta F}{A},\qquad
 \frac{\Delta M_c}{M_0}=\frac{\Delta F}{A\mu_c^2},\\
 \Delta\delta_c&=-\frac{\partial_\delta\Delta F+C\Delta\mu_c}{B}.
\end{aligned}
\]
At the three-quantum endpoint, $\mu_c=0.0641125$ and $A=-5.44988\times10^{-7}$. A mass displacement of $0.2392M_0$, the size of the last occupation refinement, corresponds at first order to $|\Delta F|\simeq5.36\times10^{-10}$. This gives a quantitative sensitivity scale for the field-order comparison. The completed response in Appendix~\ref{sec:full-field-response} exceeds that scale and places the former center--mode endpoint inside a negative-response region. Equations~\eqref{eq:quartic-field}--\eqref{eq:field-insertion-sum} specify the additional vertices and their accumulated order.

\section{Canonical separation and residual center--mode vertices}
\label{app:canonical-vertices}
The coordinate chart of Appendix~\ref{app:field-reduction} admits an explicit linear canonical separation. Let $\chi=2vK^{-1}f$, $\alpha=\chi/M_0$, and $\beta=-vD\alpha$. For each normalized moving mode define
\[
\begin{aligned}
 z_\nu&=-\langle\Phi_v',u_\nu\rangle/I_0,\\
 U_{a\nu}&=\langle e_a,u_\nu\rangle,\\
 V_{a\nu}&=\langle e_a,-i\omega_\nu u_\nu-vu_\nu'
                         -vz_\nu\Phi_v''\rangle.
\end{aligned}
\]
Here inner products are extended linearly in their second argument. The transformation, with $X=vt-L/2+\xi$, is
\begin{equation}
\begin{aligned}
 \xi&=Q+\sqrt{\hstar}\sum_\nu(z_\nu a_\nu+z_\nu^*a_\nu^\dagger),\\
 q&=\alpha p+\sqrt{\hstar}\sum_\nu(U_\nu a_\nu+U_\nu^*a_\nu^\dagger),\\
 \pi&=\beta p+\sqrt{\hstar}\sum_\nu(V_\nu a_\nu+V_\nu^*a_\nu^\dagger).
\end{aligned}
\label{eq:canonical-map}
\end{equation}
Here $Q$ is the canonically separated center coordinate, and $U_\nu,V_\nu$ are the columns of the displayed mode matrices. The total momentum fluctuation remains $p$. Equation~\eqref{eq:canonical-map} gives $[Q,p]=i\hstar$ and $[a_\nu,a_\mu^\dagger]=\delta_{\nu\mu}$, and reduces the quadratic energy to
\[
 \mathcal H_2=\frac{p^2}{2M_0}+\sum_\nu\hstar\omega_\nu a_\nu^\dagger a_\nu
\]
up to the vacuum energy. On the 192-point ring grid with 20 modes, the largest symplectic residual is $1.1\times10^{-12}$. Six independent evaluations of the quadratic form in random phase-space directions agree with its diagonal form to $1.1\times10^{-11}$.

Put $s=f^Tq$, $d_2=q^T(C-D^TD)q$, and $r=p-2vs$. The homogeneous cubic energy obtained from Eq.~\eqref{eq:moving-metric} and the field potential is
\begin{equation}
 \mathcal H_3=\frac{r(B-vd_2)}{I_0}-\frac{s r^2}{I_0^2}
                   -\frac16\int_0^L\sin\Phi_v\,\eta^3\dd y.
 \label{eq:cubic-vertices}
\end{equation}
The potential term in Eq.~\eqref{eq:cubic-vertices} transforms with the kinetic terms. Substituting Eq.~\eqref{eq:canonical-map} and subtracting the value at $p=0$ defines all cubic vertices containing the center momentum:
\begin{equation}
 \mathcal H_{3,\rm cm}=c_3p^3+p^2\mathcal W_1
                                      +p\mathcal W_2.
 \label{eq:cm-vertices}
\end{equation}
The operator $\mathcal W_1$ is linear and $\mathcal W_2$ quadratic in the mode operators. Equation~\eqref{eq:cubic-vertices} specifies their coefficients through field integrals. Weyl ordering supplies the quadratic operator; subtracting its vacuum expectation defines a fixed-reference-velocity convention. In this convention Eq.~\eqref{eq:cm-vertices} uses $:\mathcal W_2:$.

An independent check comes from the free travelling family:
\[
 c_3=\frac{E'''(P_0)}6=-\frac{M_0'(v)}{6M_0^3}.
\]
Here $E(P)$ is the energy of the free travelling profile and $P_0=P(v)$ its reference momentum. The cubic projection gives $c_3=-0.002118003878$, while differentiation of the family gives $-0.002118004033$. This comparison tests the simultaneous transformation of the kinetic and potential terms.

The number-conserving diagonal of $:\mathcal W_2:$ has a direct spectral check:
\[
 \frac{\mathcal H_{3,\mathrm{diag}}}{\hstar}
 =p\sum_\nu\frac{\partial\omega_\nu}{\partial P}
                         a_\nu^\dagger a_\nu,
 \qquad
 \frac{\partial\omega_\nu}{\partial P}
 =\frac1{M_0}\frac{\partial\omega_\nu}{\partial v}.
\]
Differentiating the ring eigenproblem at $v\pm10^{-5}$ agrees with all 20 projected diagonal coefficients to $3.4\times10^{-10}$. For mode 1, $\partial_P\omega_1=-0.0619737940$. The initial momentum width $\sigma_p=\hstar/(2\sigma_X)$ produces the free-packet phase scale $NT|\partial_P\omega_1|\sigma_p=0.03695$ at $N=18$. This phase measures the dispersion contribution to passage memory.

The source also transforms. To first order in mode and momentum fluctuations, with $X_Q=vt-L/2+Q$, it becomes
\begin{equation}
\begin{split}
 \mathcal V_{\rm int}={}&-\kappa J(X_Q)Z_1
 +\hstar\sum_\nu[h_\nu(X_Q)a_\nu^\dagger+\hc]Z_1\\
 &-\frac{\kappa}{2}\{g_\alpha(X_Q),p\}Z_1+\cdots,\\
 g_\alpha(X)&=\int_0^Lg(y+X)\frac{\partial_v\Phi_v(y)}{M_0}\dd y.
\end{split}
\label{eq:source-momentum}
\end{equation}
The identity $J'z_\nu+\sum_ag_aU_{a\nu}=\int g(y+X)u_\nu(y)\dd y$ recovers the original mode source. Equation~\eqref{eq:source-momentum} also gives a momentum-dependent source vertex. Both responses are evaluated in the same canonical coordinates.

For either perturbation $\mathcal V$, let $\psi_0$ solve the retained Hamiltonian and compute its tangent $\psi_1$ from
\begin{equation}
 i\dot\psi_1=H\psi_1+\frac{\mathcal V}{\hstar}\psi_0,
 \qquad \psi_1(0)=0.
 \label{eq:vertex-tangent}
\end{equation}
Tracing $\psi_1\psi_0^\dagger+\psi_0\psi_1^\dagger$ gives $\delta\rho$. Applying Eq.~\eqref{eq:functional} to this matrix evaluates the conditional-overlap response of Eq.~\eqref{eq:vertex-tangent}. The corresponding mixed response subtracts the tangent of the translational control; the prescribed-trajectory control has $p=0$.

\subsection{Weyl projection and vacuum contractions}
For the static field vertices, write $b=(p,a_1,\ldots,a_K,a_1^\dagger,\ldots,a_K^\dagger)$ and let $T^{(r)}$ be the symmetric coefficient tensor of degree $r$, including the factors of $\sqrt{\hstar}$ from the canonical map. A barred mode index denotes a creation operator. Before Fock projection, Weyl ordering gives
\[
\begin{aligned}
 \mathcal H_3={}&:\!\sum_{ijk}T^{(3)}_{ijk}b_ib_jb_k\!:
       +3\sum_{\nu k}T^{(3)}_{\nu\bar\nu k}b_k,\\
 \mathcal H_4={}&:\!\sum_{ijkl}T^{(4)}_{ijkl}b_ib_jb_kb_l\!:\\
 &+6\sum_{\nu ij}T^{(4)}_{\nu\bar\nu ij}:b_ib_j:
       +3\sum_{\nu\mu}T^{(4)}_{\nu\bar\nu\mu\bar\mu}.
\end{aligned}
\]
The sums over contracted indices run over all retained modes. Applying the subsequent energy and occupation projections to these operators preserves the contractions, including contributions of modes whose excited states lie above the energy bound. This construction produces the mode-count dependence resolved in Appendix~\ref{sec:full-field-response}. The center--mode coefficients reconstructed from the symmetric tensor agree with their independent expressions to below $5\times10^{-19}$ in the tested four-mode case. With the extra blocks switched off, a one-passage propagation reproduces the earlier conditional loss to $2.3\times10^{-16}$. With source coupling set to zero, the chain and environment factorize; a three-passage check gives conditional losses below $2.5\times10^{-15}$.

\section{Matching the physical spectrum and translational mass}
\label{app:physical-matching}
We use diagonal frequency counterterms and scalar momentum counterterms to define a common physical reference at each finite regulator. The matched energy Hamiltonian is
\begin{equation}
 \mathcal H^{\rm match}_K=\mathcal H_{{\rm f4},K}
 +\hstar\sum_\nu d_{\nu,K}n_\nu+c_{1,K}p+c_{2,K}p^2.
 \label{eq:matched-hamiltonian}
\end{equation}
All nonlinear vertices and source operators retain their field-derived coefficients. The target gaps are the reference frequencies, with $\omega_{1,\rm phys}T=6\pi+\delta_{\rm phys}$. The target translational curvature is $\mu/M_0$.

Let $e_{r,K}(\mathbf d)$ and $|r,K\rangle$ be the frequency eigenvalues and states of the source-free mode block at $p=0$. Its ground state is $|g_K\rangle$. Each one-particle branch is identified by its overlap with the corresponding reference one-quantum state. We determine $\mathbf d$ from
\begin{equation}
 e_{\nu,K}(\mathbf d)-e_{0,K}(\mathbf d)=\omega_{\nu,\rm phys}.
 \label{eq:matched-gaps}
\end{equation}
Newton iteration for Eq.~\eqref{eq:matched-gaps} uses the Hellmann--Feynman Jacobian
\[
 J_{\nu j}=\langle\nu,K|n_j|\nu,K\rangle
                 -\langle g_K|n_j|g_K\rangle.
\]
The branch assignments are obtained by maximizing the total one-quantum overlap. At energy bound 5, the smallest assigned squared overlap is $0.9827$ for 20 modes and $0.9674$ for 24 modes.

Write the momentum-dependent interaction energy as $A_1p+A_2p^2+A_3p^3+A_4p^4$, where $A_j$ contains the corresponding coefficients of the cubic and quartic field blocks. Differentiation of the ground-state eigenvalue gives
\begin{equation}
\begin{aligned}
 c_{1,K}&=-\langle g_K|A_1|g_K\rangle,\\
 c_{2,K}&=-\langle g_K|A_2|g_K\rangle\\
 &\quad-\sum_{r\ne0}
 \frac{|\langle r,K|A_1|g_K\rangle|^2}
 {\hstar(e_{0,K}-e_{r,K})}.
\end{aligned}
\label{eq:matched-curvature}
\end{equation}
The source-free ground-branch energy of Eq.~\eqref{eq:matched-hamiltonian} satisfies
\[
 \mathcal E'_{0,K}(0)=0,\qquad
 \mathcal E''_{0,K}(0)=\frac{\mu}{M_0}.
\]
The virtual-transition term in Eq.~\eqref{eq:matched-curvature} includes the effect of cubic momentum--mode vertices. For 20 modes at the reference phase and energy bound 5, the direct and virtual quadratic coefficients are $1.26194\times10^{-4}$ and $-7.12432\times10^{-7}$, respectively. The resulting counterterms are $(c_1,c_2)=(1.28918\times10^{-6},-1.25481\times10^{-4})$. For 24 modes they are $(2.49895\times10^{-6},-1.74043\times10^{-4})$.

\begin{figure*}[t]
\centering
\includegraphics[width=\textwidth]{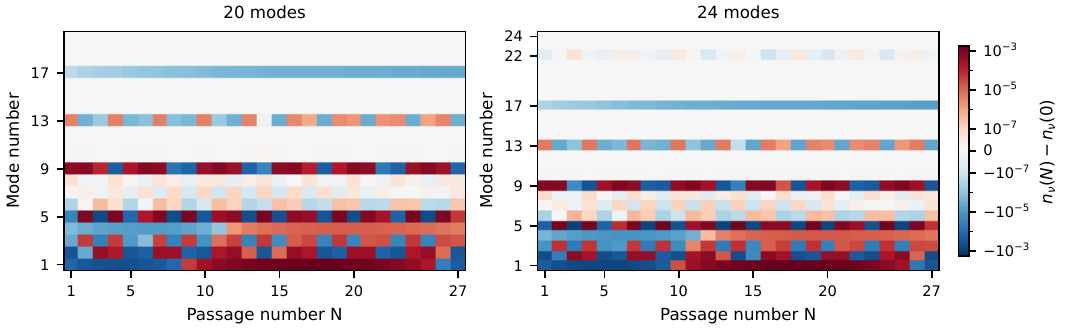}
\caption{Passage-resolved harmonic occupation increments for the matched-ground-state preparation used in Fig.~\ref{fig:mode-distribution}. Both panels use the same symmetric logarithmic color scale, with a linear interval $|n_\nu(N)-n_\nu(0)|\leq10^{-7}$.}
\label{fig:mode-passages}
\end{figure*}

All 20 or 24 one-quantum states are retained at energy bound 5. The maximum gap residual is $1.1\times10^{-14}$ in both cases. Independent ground-energy differences at momenta $\pm0.0015$ and $\pm0.00075$, followed by Richardson extrapolation, reproduce $\mu/M_0$ to $2.0\times10^{-13}$ and $9.4\times10^{-13}$, respectively. The matching is performed in the source-free system before chain propagation.

The prescribed-trajectory control uses the same matched mode block at $p=0$. The translational control has the specified free curvature $\mu/M_0$ directly. For the initial mode state we compare the reference harmonic vacuum $|0\rangle$ with $|g_K\rangle$ at zero momentum. In both preparations the initial state factorizes into the same Gaussian translational packet, the chosen mode state, and the chain state. This comparison separates the response to parameter matching from that to initial vacuum preparation.

\section{Mode moments and spectral projection checks}
\label{app:mode-diagnostics}

The mode diagnostics use the normalized joint state in the same retained Fock and translational spaces as the propagation. Lowering a retained occupation vector stays inside the energy bound. Consequently, the direct occupation sum and $\|a_\nu\Psi\|^2$ evaluate the same $n_\nu$. Across all 27 passages, the occupation identities and the conditional-distance identity agree to within $5\times10^{-14}$. The complete eigenbasis projection preserves normalization to $2\times10^{-15}$ and the conditional squared distance to $2\times10^{-18}$. The initial excitation probability in this basis is below $2\times10^{-31}$.

The normal covariance matrix is
\[
 C_{\nu\mu}=\langle a_\nu^\dagger a_\mu\rangle
              -\langle a_\nu\rangle^*\langle a_\mu\rangle.
\]
Its diagonal and positive semidefiniteness provide independent checks of the mode moments. At passage 27, $\Tr C=1.58198\times10^{-4}$ for 20 modes and $1.59352\times10^{-4}$ for 24 modes. The corresponding sums of harmonic occupations are $0.00855050$ and $0.01657733$.

The initial mean amplitudes can be estimated from the linear terms generated by the cubic vacuum contractions. Put
\[
 \ell_\nu=\hstar^{-1}\langle1_\nu|\mathcal H_3(p=0)|0\rangle.
\]
Writing $\alpha_\nu^{(0)}$ for the initial ground-state mean $\langle a_\nu\rangle$, the leading displacement estimate is
\begin{equation}
 \alpha_\nu^{(0)}\simeq-\frac{\ell_\nu}{\omega_{\nu,\rm phys}}.
 \label{eq:vacuum-displacement}
\end{equation}
Relative to the vector of exact matched-ground-state means, Eq.~\eqref{eq:vacuum-displacement} has Euclidean relative error $0.108\%$ for 20 modes and $0.155\%$ for 24 modes. The contracted sums over retained modes change these linear coefficients and hence the displacement of the low modes.

Figure~\ref{fig:mode-passages} shows the signed increments $n_\nu(N)-n_\nu(0)$ at each passage. The occupations evolve about the deformed reference state: after 27 passages their sum has decreased by $6.96\%$ and $5.67\%$ for the two mode counts. The eigenbranch probabilities in Fig.~\ref{fig:mode-distribution} resolve excitation relative to that reference state. The conditional means and covariance matrix retain the phase information of the mode amplitudes.

\end{document}